\documentclass[12pt]{article}

\usepackage[english]{babel}

\usepackage[nottoc,notlot,notlof]{tocbibind}
\usepackage{amssymb,mathrsfs,amsfonts}
\usepackage{amsmath, amsthm, xpatch, dsfont,ushort,sgamevar, float, framed, multirow, array, makecell, comment, url, graphicx, subcaption, mathtools, fnpct, setspace, pifont,bm,bbm, accents}

\usepackage{nicefrac, xfrac}

\usepackage{epstopdf}

\newcommand{\ubar}[1]{\underaccent{\bar}{#1}}

\usepackage{lmodern}
\usepackage[final, babel]{microtype}

\DeclarePairedDelimiterX{\infdivx}[2]{(}{)}{%
#1\;\delimsize\|\;#2%
}

\usepackage{lipsum}
\usepackage[utf8]{inputenc} 
\usepackage[T1]{fontenc} 
\usepackage{titlesec}

\usepackage[round,comma,authoryear,longnamesfirst,sort&compress]{natbib}
\usepackage[bottom, multiple]{footmisc}

\usepackage[top=1in, bottom=1.25in, left=1in, right=1in]{geometry}
\usepackage{xcolor}
\definecolor{navyblue}{rgb}{0.0, 0.13, 0.5}
\definecolor{amaranth}{rgb}{0.9, 0.17, 0.31}
\definecolor{brightpink}{rgb}{1.0, 0.0, 0.5}
\definecolor{cgreen}{rgb}{0.0, 0.42, 0.24}
\definecolor{cobalt}{rgb}{0.0, 0.28, 0.67}
\definecolor{coquelicot}{rgb}{1.0, 0.22, 0.0}
\definecolor{dukeblue}{rgb}{0.0, 0.0, 0.61}

\usepackage[pagebackref=true]{hyperref}
\hypersetup{
colorlinks=true,
linkcolor=black,
filecolor=black,
citecolor=cobalt,      
urlcolor=cobalt,
}
\usepackage[capitalize,noabbrev,nameinlink]{cleveref}

\renewcommand*{\backrefalt}[4]{%
\ifcase #1 %
No citations.%
\or
(#2).%
\else
(#2).%
\fi
}

\DeclareMathAlphabet{\mathsf}{OT1}{\sfdefault}{m}{n}
\SetMathAlphabet{\mathsf}{bold}{OT1}{\sfdefault}{b}{n}

\usepackage{enumitem}

\newcommand*\diff{\mathop{}\!\mathrm{d}}
\usepackage{tocloft}

\newtheorem{bprop*}{Proposition}
\newtheorem{prop*}{Proposition}[]
\newtheorem{thm*}{Theorem}[]
\newtheorem{claim*}{Claim}[]
\newtheorem{cor*}{Corollary}[]
\newtheorem{ax*}{Axiom}[]
\newtheorem{lem*}{Lemma}[]
\newtheorem{rem*}{Remark}[]
\newtheorem{asp*}{Assumption}[]
\newtheorem{pty*}{Property}[]
\newtheorem{def*}{Definition}[]
\newtheorem{cond*}{Condition}[]

\newtheorem*{propnonumber*}{Proposition}

\usepackage{caption}
\usepackage[labelfont=bf]{caption}

\usepackage{tikz}
\usepackage{pgfplots}
\usetikzlibrary{automata,positioning}
\usetikzlibrary{arrows,intersections}

\newtheorem{examplex}{Example}
\newenvironment{eg*}
{\pushQED{\qed}\examplex}
{\popQED\endexamplex}

\newcommand{\ra}{\rightarrow}

\newcommand{\ve}{\varepsilon}

\newcommand{\bP}{{\bf P}}

\newcommand{\bE}{\mathbf{E}}
\newcommand{\bR}{\mathbf{R}}

\newcommand{\expo}{ \operatorname{e}}
\newcommand{\gl}{\lambda}
\newcommand{\ga}{\alpha}
\newcommand{\gb}{\beta}
\newcommand{\gd}{\delta}
\newcommand{\gs}{\sigma}

\newcommand{\gk}{\kappa}
\newcommand{\egg}{e.g.,~}
\newcommand{\ie}{i.e.,~}

\usepackage[toc]{appendix}

\renewcommand{\texttt}[1]{%
\begingroup
\ttfamily
\begingroup\lccode`~=`/\lowercase{\endgroup\def~}{/\discretionary{}{}{}}%
\begingroup\lccode`~=`[\lowercase{\endgroup\def~}{[\discretionary{}{}{}}%
\begingroup\lccode`~=`.\lowercase{\endgroup\def~}{.\discretionary{}{}{}}%
\catcode`/=\active\catcode`[=\active\catcode`.=\active
\scantokens{#1\noexpand}%
\endgroup
}

\makeatletter
\renewenvironment{proof}[1][\proofname]{%
\par\pushQED{\qed}\normalfont%
\topsep6\p@\@plus6\p@\relax
\trivlist\item[\hskip\labelsep\bfseries#1\@addpunct{.}]%
\ignorespaces
}{%
\popQED\endtrivlist\@endpefalse
}
\makeatother

\usepackage{soul}

\pgfplotsset{compat=1.14}
\patchcmd{\quote}{\rightmargin}{\leftmargin 4em \rightmargin}{}{}

\makeatletter
\AddToHook{cmd/appendix/before}{\def\cref@section@alias{appendix}\def\cref@subsection@alias{appendix}}
\makeatother

\begin{document}

\title{Reputational cheap talk:\\[-0.4em] influentialness and welfare\thanks{I am grateful to Johannes H\"orner, Larry Samuelson, and Marina Halac for their support. I also thank Marco Ottaviani for helpful discussion. I have also benefited from conversations with Ian Ball, Heski Bar-Isaac, Dirk Bergemann, Simon Board, Jimmy Chan, Joyee Deb, Florian Ederer, Mira Frick, Jan Kn\"{o}pfle, Tan Gan, Axel Gautier, John Geanakoplos, Ryota Iijima, Chuqing Jin, Navin Kartik, R.\ Vijay Krishna, Patrick Lahr, Bart Lipman, Melody Lo, Chiara Margaria, Erik Madsen, Michael Manove, Weicheng Min, Heinrich Nax, Aniko \"{O}ry, Harry Pei, Yujie Qian, Anna Sanktjohanser, Sara Shahanaghi, Eran Shmaya, Wing Suen, Egor Starkov, Philipp Strack, Yiman Sun, Juuso V\"alim\"aki, Alex Wolitzky, Asher Wolinsky, Kai Hao Yang, Weijie Zhong, Zhen Zhou, John Zhu, and the audiences at AMES 2021, Boston University, CMES 2021, CityU of Hong Kong Microeconomics Conference 2025, CUHK, ESAM 2021, HKBU, SAET 2025, SIOE 2021, LED YES, Mannheim virtual IO seminar, NASMES 2021, NTU, NUS, PKU-NUS Conference 2021, Tsinghua BEAT 2021, and Yale.}
}
\author{Allen Vong
\thanks{National University of Singapore. Email: \href{mailto:allenvongecon@gmail.com}{\texttt{allenv@nus.edu.sg}}.}
}
\date{\today}
\maketitle

\vspace*{-15px}
\begin{abstract}
A sender communicates private information about a hidden state to a receiver who seeks to match his action to that state. The sender strives to appear informed at the receiver's expense. I characterize informative equilibria under a broad class of signal structures and show that, when they exist, they are essentially unique. I show that informative equilibria can be noninfluential, in which case the receiver does not benefit from communication and relies only on prior information. My main results identify a complementarity that sufficiently precise prior information helps restore influential communication and characterize how the receiver's welfare depends on the quality of prior information. I also characterize how the sender's initial reputation for being informed shapes these results.


\medskip


\end{abstract}

\thispagestyle{empty}
\pagebreak
\setcounter{page}{1}

\section{Introduction}
\label{sec:intro}


Decision makers often face a tradeoff between relying on expert advice and relying on prior information. For example, fund managers may weigh analysts' recommendations to sell widely praised assets, politicians may weigh advisors' warnings against popular policies, and patients may weigh doctors' recommendations against popular treatments.

This tradeoff is especially salient when advisors primarily care about labor-market perceptions of their ability rather than their advisees' welfare. The reputational cheap talk literature, pioneered by \citet{trueman1994analyst} and \citet{ottaviani2006professional,ottaviani2006reputational,ottaviani2006strategy}, has shown that expert advice can be informative in these settings, but does not address this tradeoff. Informative communication affects beliefs but need not affect decisions.

In this paper I study this tradeoff and its allocative consequences. I study a model where a sender faces a receiver who wants to match his action with a binary hidden state. The sender privately knows her type, high or low, and observes a private signal from a statistical experiment about the state; the low type's experiment is noisier than the high type's. The sender sends a report of the state to the receiver who then takes an action. A labor market evaluates the sender's value given her report and the eventual revelation of the state, assigning a higher value to the high type. The sender maximizes her reputation, namely the market's belief that she is a high type, and thus her market value. I study all informative, \ie non-babbling, equilibria and examine whether they are influential, namely whether the receiver matches his action with the sender's report. If an equilibrium fails to be influential, communication is outcome-irrelevant: the receiver takes the ex ante optimal action irrespective of the report and obtains his reservation payoff.

I first characterize the existence and structure of informative equilibria and show that, when they exist, they are essentially unique, namely unique up to outcome equivalence. In the informative equilibrium, each sender type trades off the probability that each report is correct against the distribution of reputations the report induces. She may send a misleading report, namely a report of the state she deems unlikely, for a favorable distribution of reputations. For some experiments and sufficiently noisy prior state beliefs, only the conformist report, namely a report of the ex ante likely state, may be misleading; otherwise, only the contrarian report, namely a report of the ex ante unlikely state, may be misleading. The receiver matches his action with a conformist report. Therefore, the equilibrium is influential if and only if he also matches his action with a contrarian report. For a nondegenerate set of parameters and experiments valuable to the receiver, the informative equilibrium is noninfluential: the receiver views the contrarian report as too likely to be misleading.



My first main result shows that if the prior state belief is sufficiently precise, then the informative equilibrium is influential provided that the high type's experiment is sufficiently informative. Intuitively, in such a setting, the high type sends a contrarian report only when her signal strongly favors the ex ante unlikely state. The low type perceives this event as rare and seldom issues a contrarian report in order to mimic the high type. The receiver therefore views the contrarian report as rarely misleading, and so the equilibrium is influential.

This result identifies a novel complementarity between high-quality prior information and reliable expertise. It contrasts with the conventional view that prior information crowds out valuable private information (\egg \citealp{morris2002social}; \citealp{chen2012value}), and helps explain well-known episodes in which contrarian advice shaped decisions. For instance, Andy Grove urged Intel to exit its memory-chip business and focus on microprocessors; although controversial at the time, Intel adopted this strategy, which proved key to the firm's success \citep{grove1999only}. A similar pattern appears in George Kennan's ``Long Telegram'' which challenged the Yalta Axioms and shaped the United States' Cold War strategy \citep{gaddis2012george}. 

This complementarity also contrasts with the conventional wisdom that decision makers substitute expertise for noisy public information. I show that this substitution is also present in my model: when the prior state belief is sufficiently noisy, the informative equilibrium is influential because his reservation payoff is low. Consequently, provided that the high type's experiment is sufficiently informative, both sufficiently noisy and sufficiently precise prior information help sustain the equilibrium's influentialness, but via different mechanisms that have fundamentally different payoff implications. The receiver fares better when relying on prior information than on the sender if, in the latter setting, prior information is sufficiently noisy that he relies on the sender despite the risk of being misled, reflecting the substitution mechanism. He fares worse if, in the latter setting, prior information is sufficiently precise that contrarian reports are rarely misleading, reflecting the complementarity mechanism.






My second main result shows that, when the high type's experiment is sufficiently informative, the receiver's informative-equilibrium payoff, which measures allocative efficiency, is essentially single-dipped in the prior state belief. Intuitively, more precise prior information raises the probability that the low type sends a conformist report. A marginal increase in the prior state belief then harms the receiver when this belief is sufficiently noisy and the conformist report may be misleading. Otherwise, only the contrarian report may be misleading, so a higher prior state belief benefits the receiver, even if it lies in a range that eliminates the equilibrium's influentialness.

Prior state beliefs are often interpreted as reflecting public information about the state.\footnote{They may be viewed as interim beliefs induced by public signals in an extended setting.} My results provide both a caution and a rationale for efforts to improve public information, such as regulations against misinformation (\egg \citealp{funke2021guide}) and stress testing (\egg \citealp{hirtle2015supervisory}). When experts are inefficiently conformist, such efforts harm allocative efficiency unless implemented sufficiently aggressively. In contrast, when experts are inefficiently contrarian, these efforts unambiguously improve allocative efficiency and may or may not crowd out expertise. If sufficiently strong, they sustain influential expertise that outperforms the improved public information. These insights may be particularly relevant in financial markets, where career concerns often induce contrarian behavior (\egg \citealp{zitzewitz2001measuring}; \citealp{bernhardt2006herds}; \citealp{bozanic2019analyst}).







My results also offer a screening-based perspective on why highly skilled experts benefit from information technologies such as artificial intelligence, complementing the existing view that these technologies increase experts' productivity (see, \egg \citealp{ide2025artificial}). The requirement in my first main result that the high type's experiment be sufficiently informative shows that these technologies crowd out mediocre experts but not highly skilled ones.


Finally, I characterize how the sender's initial reputation shapes the effects of prior information on influentialness and allocative efficiency. I show that when the high type's experiment is sufficiently informative, the range of prior state beliefs supporting influentialness strictly expands with the sender's initial reputation, because a more reputable sender has weaker incentives to issue a misleading report for reputation gains. Nonetheless, a higher initial reputation only weakly improves the receiver's payoff. This improvement is strict if and only if the resulting equilibrium is influential, in which case the reduced likelihood of a misleading report is outcome-relevant.

These results further inform institutional design for allocative efficiency. When experts are inefficiently conformist, reputation-enhancing policies, such as stricter certification or occupational licensing, dominate public-information improvements unless the latter are sufficiently strong. When experts are inefficiently contrarian, reputation improvements are generally less effective than public-information improvements because they need not improve welfare, but they complement the latter by expanding the range of information environments that sustain influential expertise.


I close the paper by extending the analysis to several variants of the model relevant for different applications. I consider a setting in which the sender's message is hidden from the market, so that the market instead infers her type from the receiver's action. I also study environments in which the receiver chooses whether to delegate decision-making to the sender, or whether to obtain a message from her. Finally, while my main analysis follows \citet{trueman1994analyst} and \citet[Section 6]{ottaviani2006reputational} in assuming that the sender privately knows her type, I consider a variant in which she, like the receiver, is uncertain about it.

\paragraph{Related literature.} 
This paper provides the first characterization of influentialness and allocative efficiency of informative equilibria in reputational cheap talk. Prior work has abstracted from the receiver's decision making and payoff, focusing on the informativeness of reputational cheap talk; see \citet[Section 5.1.3]{bergemann2021information} for a recent overview.\footnote{ Exceptions that incorporate decision making in response to reputational cheap talk do not study influentialness and its payoff implications. For example, the voting models of \citet{levy2007decision, levy2007decision2} and \citet{visser2007committees} assume commitment to a decision rule, and the two-sender sequential model of \citet{bag2019sequential} builds on a stage game where an influential equilibrium always exists.} My analysis also makes a technical contribution. To highlight the flexibility of my results, I study a broad class of nonparametric experiments, obtaining a new characterization of informative equilibrium that generalizes existing results, which has largely focused on parametric experiments; see \citet{ottaviani2006professional, ottaviani2006reputational, ottaviani2006strategy} for examples.\footnote{While \citet{ottaviani2006reputational} allow for nonparametric experiments to show that truthful reporting fails to be an equilibrium, their equilibrium constructions rely on parametric specifications.} \cite{shahanaghi2024investment} studies a related reputational timing game in which the high type has a nonparametric experiment but the low type is perfectly uninformative.

My results show that influentialness in reputational cheap talk differs from that in partisan cheap talk \`a la \citet{crawford1982strategic}. In the latter, an influential equilibrium exists if and only if the exogenous difference between the sender's and receiver's preferred actions is sufficiently small (see, \egg \citealp{sobel2013giving}).\footnote{An exception is \citet{chakraborty2010persuasion}, who show that influential equilibria may exist under a multidimensional state space even when this difference is arbitrarily large.} In my model, an informative equilibrium is influential if and only if the probability of a misleading contrarian report, endogenously determined in the equilibrium, is sufficiently small. My results also contrast with \citet{chen2012value} who shows that high-quality public information yields uninformative and noninfluential communication in partisan cheap talk.\footnote{Other partisan cheap talk models study the effects of prior information when the sender is perfectly informed, but are less related to my setting because prior information does not affect the sender's state belief in those models (\egg \citealp{chen2015information}).}

My results contrast with bad-reputation models, where equilibrium is typically uninformative and all equilibria are noninfluential (\citealp{morris2001political}; \citealp{kartik2019informative}). Because my results also apply to delegation, they contrast with bad-reputation delegation models as well (\citealp{ely2003bad}; \citealp{min2025bad}). They also differ from good-reputation models, where influential communication arises because strategic senders pool with an informed commitment type who acts in the receiver's interest (\citealp{sobel1985theory}; \citealp{benabou1992using}); my model has no commitment type.

\section{Model}
\label{sec:model}

There are a sender (she) and a receiver (he). The sender has a private type $t \in \{ h, l\}$, high ($h$) or low ($l$). There is a hidden state $s \in S :=\{0,1\}$. 

Nature first draws type $t$ to be high with probability $p \in (0,1)$ and low otherwise, and independently draws state $s$ to be 1 with probability $\mu \in  [\frac 1 2,1)$ and 0 otherwise; it is without loss of generality that state 1 is ex ante more likely than state 0 to be true. The sender receives a private signal $\ell \in [0, \infty]$,\footnote{I use the extended half-line as the set of signals because, as will be explained momentarily, I label signals directly as their likelihood ratios.} and sends the receiver a message $m \in S$, interpreted as a report of the state.\footnote{Therefore, communication is assumed to be binary, following prior work in the literature. Whether binary messages entail loss of generality is an open question in the literature; an exception is \cite{ottaviani2006professional}, who show that binary communication is without loss of generality under a multiplicative linear experiment when the sender does not know her type.} Because state 1 is ex ante more likely, I call report 1 a conformist report and report 0 a contrarian report. Finally, the receiver takes an action $a \in S$.\footnote{My results extend if the receiver observes an independent private signal about the state, as in \cite{lai2014expert}, so long as this signal is sufficiently noisy relative to both sender types' signals.} 

Conditional on type $t$ and state $s$, the sender's signal $\ell$ is drawn from a continuously differentiable distribution function $F(\cdot|t,s)$ with density $f(\cdot|t,s)$ and support $L(t) := [\ubar \ell^t, \bar \ell^t] \subseteq [0, \infty]$, where $\ubar \ell^t < \bar \ell^t$. The support is state-independent so that signals are noisy. Signals are labeled as likelihood ratios, $\ell = f(\ell|t,1)/f(\ell|t,0)$.\footnote{\label{fn:lr}This labeling is a modeling shortcut and entails no loss of generality, so long as the focus is on experiments that satisfy the monotone likelihood ratio property. See \cref{fn:g} in \cref{eg:mle} for an illustration.} The monotone likelihood ratio property holds: each type deems the state more likely to be 1 given a higher signal.  I refer to $F^t:=(F(\cdot|t,s))_{s \in S}$ as type-$t$ sender's experiment and $F := (F^h, F^l)$ as a profile of experiments. I refer to the tuple $\xi := (\mu, p, F)$ as an information structure, and impose the following assumption on $F$ throughout.

\begin{asp*}[The low type's experiment is noisier than the high type's]\label{asp:lr2} It holds that:
\begin{itemize}\itemsep0em
\item[\textnormal{(a)}] $\ubar \ell^h < \ubar \ell^l$ and $\bar \ell^l < \bar \ell^h$.
\item[\textnormal{(b)}] For each state $s$ and each likelihood ratio $\ell \in (\ubar \ell^l, \bar \ell^l)$,
\begin{align}\label{eq:hrates}
\frac{f(\ell|l,s)}{F(\ell|l,s)} > \frac{f(\ell|h,s)}{F(\ell|h,s)} \quad \text{ and } \quad \frac{f(\ell|l,s)}{1-F(\ell|l,s)} > \frac{f(\ell|h,s)}{1-F(\ell|h,s)}.
\end{align}
\item[\textnormal{(c)}] $\mu<1/(1+\ubar \ell^h)$.
\end{itemize}
\end{asp*}

\cref{asp:lr2} enables a tractable equilibrium characterization. Part (a) states that the high type can form more extreme (and thus more informative) likelihood ratios than the low type. Part (b) states that both the reverse hazard rate and the hazard rate of the low type are higher than those of the high type; the low type is thus more likely than the high type to form mediocre (and thus noisy) likelihood ratios. Part (c) states that the high type's experiment can yield signals given which she believes state 0 is more likely.\footnote{As $\ubar \ell^h = f(\ubar \ell^h|h,1)/f(\ubar \ell^h|h,0)$, (c) is equivalent to $\mu f(\ubar \ell^h|h,1)/[\mu f(\ubar \ell^h|h,1) + (1-\mu)f(\ubar \ell^l|h,0)] < \frac 1 2$.} As will be clear, this ensures that the receiver's decision problem is nontrivial. \cref{asp:lr2} implies that the high type's experiment Blackwell-dominates the low type's, but the converse is not true because Blackwell dominance requires neither that the support of the low type's experiment be a strict subset of that of the high type's nor that \eqref{eq:hrates} hold for all likelihood ratios.

A familiar example that satisfies \cref{asp:lr2} is:

\begin{eg*}\label{eg:mle} \normalfont
For each type $t$, let $x(t) \in (0,1]$, with $x(l) < x(h)=1$. In state $s$, this sender type's signal $\ell$ is drawn according to the density
\begin{align}\label{eq:densityF}
f(\ell|t,s) = ((1-s)+s \ell) \dfrac{2}{x(t)(1+\ell)^3},
\end{align}
with support
\begin{align*}
L(t)=\left[\frac{1-x(t)}{1+x(t)},\frac{1+x(t)}{1-x(t)}\right].
\end{align*}
This is \citeauthor{ottaviani2006professional}'s (\citeyear{ottaviani2006professional}) multiplicative linear experiment, with signals labeled as likelihood ratios.\footnote{\label{fn:g}To see this, for each type $t$ and state $s$, let $g(\cdot|t,s):[0,1] \ra \bR$ be given by
\begin{align}
g(y|t,s) &= x(t) \underbrace{2[ (1-s)(1-y) + s y ]
}_{\textnormal{informative component}}
+ ~(1-x(t)) \cdot \hspace*{-40px} \underbrace{1 \vphantom{[ (1-s) k(1-y)^{k-1} + s k y^{k-1} ]}
}_{\textnormal{uninformative component}} \hspace*{-30px},
\label{eq:gmledensitym}
\end{align}
and define $\gl(y|t) := g(y|t,1)/g(y|t,0)$. \eqref{eq:gmledensitym} is a density of a signal $y$ on $[0,1]$ that is not labeled as a likelihood ratio, with the high type's signal more likely to be drawn from the informative component than from the uninformative one. The function $\gl(y|t)$ is the likelihood ratio at signal $y$, which is monotone in $y$ so that $\gl^{-1}(\cdot|t)$ is well-defined, and \eqref{eq:densityF} gives the density of this likelihood ratio. To derive \eqref{eq:densityF}, let $G(\cdot|t,s)$ denote the cumulative distribution function corresponding to \eqref{eq:gmledensitym}. The distribution $F(\ell|t,s)$ of the resulting likelihood ratio $\ell$ is given by the composite function $F(\ell|t,s) = G(\gl^{-1}(\ell|t)|t,s)$, and differentiating yields \eqref{eq:densityF}.} In \cref{sec:multiplicativelinearappendix}, I show that \eqref{eq:densityF} satisfies \cref{asp:lr2}(a) and (b); \cref{asp:lr2}(c) holds because the required inequality simplifies to $\mu < 1$.
\end{eg*}











I do not restrict attention to experiments that are symmetric across the states, although the literature typically focuses on these experiments, such as the multiplicative linear experiment. Formally, for each type $t$, an experiment $F^t$ is symmetric (across the states) if for each $\ell \in [\ubar \ell^t,\bar \ell^t]$, $F(\ell | t,0)=1-F(1/\ell | t,1)$.\footnote{\label{fn:symmetry}For any signal distribution $G(y|t,s)$ with support normalized to $[0,1]$ and signals $y$ not labeled as likelihood ratios, as in \cref{eg:mle}, symmetry takes the more familiar form that the state-1 distribution is the mirror image of the state-0 distribution, $G(y|t,0) = 1-G(1-y|t,1)$. In \cref{sec:fnsymmetry} I show that, under the monotone likelihood ratio property, this is equivalent to $F(\ell|t,0) = 1 -F(1/\ell|t,1)$.}

Let $F^* := ( F^*(\cdot|s) )_{s \in S}$ denote a perfect experiment that fully reveals the state: $F^*(\cdot|s)$ is a distribution on $[0,\infty]$, assigning a unit mass to $\{0\}$ if $s=0$ and a unit mass on $\{\infty\}$ if $s=1$. Let $\diff(\cdot, \cdot)$ denote the Prokhorov metric. Even though the high type's experiment $F^h$ cannot be equal to $F^*$ by assumption, as $F^*(\cdot|s)$ is singular for each $s$, $F^h$ can be arbitrarily close to $F^*$ with respect to the Prokhorov metric, namely $\max_{s \in S} \diff(F(\cdot|h,s), F^*(\cdot|s))$ can be arbitrarily close to zero. For any $\mu \in [\frac 1 2, 1)$, \cref{asp:lr2}(c) holds when $F^h$ is sufficiently close to $F^*$, because then $\ubar \ell^h$ is sufficiently close to zero.

My main results require that the high type's experiment is sufficiently informative, namely sufficiently close to $F^*$, such as in \cref{eg:hyper} below, but unlike the multiplicative linear experiment in \cref{eg:mle}. 

\begin{eg*}\label{eg:hyper}\normalfont
The low type's experiment is given by the multiplicative linear experiment in \cref{eg:mle}, with $0<x(l)<1$. In each state $s$, the high type's signal is drawn from a simple hyperexponential distribution $F^k(\cdot|h,s)$ with density
\begin{align*}
f^k(\ell|h,s)
= (1-s+s \ell) \!\left[ \!\left( 1- \frac 1 k\right)\! k \expo^{-k\ell}
+
\frac 1 k \!\left( \frac{k}{k^2-k+1} 
\expo^{- k \ell / (k^2-k+1)} \right)\! \right]\!,
\end{align*}
and support $[0,\infty]$, for some positive integer $k$; the experiment becomes Blackwell more informative given a higher $k$. In \cref{sec:hyperexponentialappendix}, I prove the following properties of this high type's experiment that will be helpful for interpreting my main results. For sufficiently large $k$, \cref{asp:lr2} holds.\footnote{It need not hold for all $k$, depending on the value of $x(l)$. While alternative specifications of the high type's experiment can ensure that \cref{asp:lr2} holds for every $k$, as well as the weak convergence to $F^*$, the present example suffices for my main results.} Moreover, $F^k$ converges weakly to $F^*$.\footnote{Recall that the Prokhorov metric metrizes the topology of weak convergence of measures (see, \egg \citealp[Theorem 6.8]{billingsley2013convergence}).} 
\end{eg*}


Type-$t$ sender's strategy is a measurable mapping $\gs^t: L(t) \ra \Delta(S)$ from her signal to a distribution over reports; $\gs^t(m|\ell)$ denotes the probability that she reports $m$ given signal $\ell$. Let $\gs=(\gs^h, \gs^l)$ be a profile of both types' strategies. The receiver's strategy is a mapping $\tau: S \ra \Delta(S)$ from each report to a distribution over actions.


Each sender type concerns maximizing her value to an external market given the market's observation of her report and the eventual revelation of the state.\footnote{The receiver's action is irrelevant for the market's valuation of the sender. On the other hand, as is standard, eventual revelation of the state allows for informative communication to possibly be transmitted in equilibrium; see \citet[Footnote 20]{bergemann2021information} for further discussion. My main results extend so long as the state is eventually revealed to the market with sufficiently high probability, or if a sufficiently precise signal about the state is eventually revealed to the market.} Let $v(t)$ denote the value of type-$t$ sender to the market, with $v(h) > v(l)$ so that the high type is more valuable than the low type. Given report $m$, revelation of state $s$, and the market's conjecture $\hat \gs = (\hat \gs^h, \hat \gs^l)$ of the sender's strategy, I denote by $\rho(t|m, s, \hat \gs)$ the probability that the market assigns to the sender being type $t$ and, following the literature, define each sender type's payoff as her expected market value:
\begin{align}\label{eq:reppayoff}
r(m, s, \hat \gs) := \rho(h|m, s,\hat \gs) v(h) + \rho(l|m, s,\hat \gs) v(l).
\end{align}
I impose the normalization $v(h)=1$ and $v(l)=0$ so that the payoff \eqref{eq:reppayoff} is simply the market's belief $\rho(h|m,s,\hat \gs)$ that the sender is a high type, which I interpret as the sender's reputation. The receiver concerns matching his action with the state. His realized payoff is $u(a,s)$, normalized to be $1$ if $a=s$ and $0$ otherwise; my results extend if the symmetry of this payoff across the states is relaxed.

The solution concept I use is Bayesian Nash equilibrium, henceforth equilibrium. In informative equilibria that I focus on, described momentarily, the sender has no observable deviation, and so focusing on Bayesian Nash equilibria, rather than perfect Bayesian equilibria or sequential equilibria, is innocuous.

\begin{def*}[Equilibrium]\label{def:eqm}
An equilibrium is a strategy profile $(\gs, \tau)$ where:
\begin{enumerate}\itemsep0em
\item Type-$t$ sender's strategy $\gs^t$ solves $\max_{\tilde \gs^t} \bE^{\tilde \gs^t} [ r(m, s, \gs) ]$ given the receiver's conjecture $\gs$, where the expectation is taken with respect to the distribution of $(m,s)$ induced by $\tilde \gs^t$.
\item The receiver's strategy $\tau$, given his conjecture $\gs$, solves $\max_{\tilde \tau} \bE^{(\gs, \tilde \tau)} \!\left[ u(a,s) \right]$, where the expectation is taken with respect to the distribution of $(a,s)$ induced by $(\gs, \tilde \tau)$.
\end{enumerate}
\end{def*}

\cref{def:eqm} is standard. Part 1 is a fixed-point condition: the equilibrium strategy $\gs^t$ is type-$t$ sender's best reply when the market conjectures that the sender plays $\gs$. Part 2 states that the receiver maximizes his payoff given his conjecture $\gs$.

If the receiver must take an action without observing a report, he optimally takes action 1 because $\mu \ge \frac 1 2$. His reservation payoff is therefore $\mu$. Accordingly, I interpret a higher $\mu$ as reflecting a more informative prior state belief.

An equilibrium exists: a babbling equilibrium exists where the sender conveys no information about her signal and the receiver takes action $1$ irrespective of the sender's message. I focus on informative, \ie non-babbling, equilibria:

\begin{def*}[Informativeness]\label{def:informativeness}
An equilibrium $(\gs, \tau)$ is informative (about the state) if in this equilibrium, there is a report $m$ sent on path, given which the receiver's belief that the state is 1 differs from $\mu$.
\end{def*}

In any informative equilibrium, both reports are sent on path. Hence, for each report $m$ and state $s$, the sender's payoff in \eqref{eq:reppayoff} is determined by Bayes' rule. Let $\neg s$ denote the state that is not $s$.

\begin{lem*}\label{lem:infthenrep}
In any informative equilibrium $(\gs, \tau)$, one of the following holds:
\begin{align}\label{eq:correctreport}
r(s, s,\gs) &> r(\neg s, s, \gs) \quad \text{ for each } s,\\
\label{eq:correctreport2}
r(s, s,\gs) &< r(\neg s, s, \gs) \quad \text{ for each } s.
\end{align}
\end{lem*}

Proofs of all formal results are in the Appendix. In any informative equilibrium, reports are informative about the sender's type because they are informative about the sender's signal and the high type's experiment is more informative than the low type's. Because the meaning of messages arises in equilibrium, either the sender's reputation is strictly higher following a correct report---one matching the state---than an incorrect one---one not matching the state, in which case \eqref{eq:correctreport} holds, or the opposite is true, in which case \eqref{eq:correctreport2} holds. Without loss of generality, hereafter I assume that \eqref{eq:correctreport} holds, which can be ensured by relabeling messages and is consistent with the interpretation of messages as state reports.

Communication is outcome-relevant in an informative equilibrium if the receiver's payoff strictly exceeds his reservation payoff.\footnote{The focus on informative equilibrium is without loss of generality. If communication is outcome-relevant in an equilibrium, then the equilibrium must be informative.} In any such equilibrium, some report on path must give the receiver a strict incentive to take the ex ante suboptimal action 0. Accordingly, throughout I refer to informative equilibria with outcome-relevant communication as influential equilibria:

\begin{def*}[Influentialness]\label{def:vcomm}
An equilibrium $(\gs, \tau)$ is influential if $\bE^{(\gs, \tau)}[u(a,s)] > \mu$ and is noninfluential otherwise.
\end{def*}

The partisan cheap talk literature sometimes adopts a weaker definition than \cref{def:vcomm}, under which influential equilibria also include those where the receiver takes action 0 only when he is indifferent between the two actions (see, \egg \citealp{sobel2013giving} for a survey). In such equilibria, the receiver obtains his reservation payoff and communication is outcome-irrelevant; \cref{def:vcomm} treats them as noninfluential.

\cref{asp:lr2}(c) ensures that the analysis of influential equilibria is nontrivial. If it fails, then the high type never has a state belief favoring state 0. \cref{asp:lr2}(a) then implies that the same is true for the low type. Then, in any equilibrium, the receiver takes action 1 irrespective of the sender's report and the equilibrium is noninfluential.

For analytical convenience, I adopt a belief-based approach to characterize equilibria, working directly with the sender's state belief induced by her signal.\footnote{This approach was first used by \cite{aumann1995repeated} in repeated games and is now widely applied in partisan cheap talk and Bayesian persuasion (see, \egg \citealp{lipnowski2020cheap} and references therein). Unlike those literatures, where the belief-based approach centers on the distribution of the receiver's state beliefs, in this paper the sender's state belief plays a key role.} Throughout, when no ambiguity arises, a state belief refers to a belief that the state is 1. Given prior state belief $\mu$ and experiment profile $F$, conditional on type $t$ and state $s$, let the distribution function of the sender's state belief induced by her signal be denoted by
\begin{align}\label{eq:HF}
H_{\mu, F}(\gb|t,s) 
= F\!\left(
\frac{1-\mu}{\mu} \frac{\gb}{1-\gb}
\middle| t, s
\right)\!, \qquad \gb \in [0,1].
\end{align}
For each type $t$, let $\ubar \gb^t$ be her lowest state belief and let $\bar \gb^t$ be her highest state belief. By \cref{asp:lr2} and the monotone likelihood ratio property, $\ubar \gb^t$ (resp., $\bar \gb^t$) is induced by signal $\ubar \ell^t$ (resp., $\bar \ell^t$), and $0 < \ubar \gb^h < \ubar \gb^l < \bar \gb^l < \bar \gb^h < 1$.

\cref{lem:signals} below is a useful preliminary result, showing that the state belief distribution in each state exhibits a single-crossing property across the two types. Consequently, the distribution of the high type's state beliefs is a mean-preserving spread of that of the low type's state beliefs \citep{rothschild1971increasing,rothschild1971increasing2}.

\begin{lem*}\label{lem:signals}
For any $\mu \in [\frac 1 2, 1)$ and $F$ satisfying \cref{asp:lr2}, for each state $s$, there is a unique $\gb^\dagger_s \in (\ubar \gb^l, \bar \gb^l)$ such that for each $\gb \in (\ubar \gb^h, \bar \gb^h)$,
\begin{align*}
H_{\mu, F}(\gb|h,s) - H_{\mu, F}(\gb|l,s) \begin{cases}
~ > 0, \qquad &\text{ if } \gb < \gb^\dagger_s,\\
~ = 0, \qquad &\text{ if } \gb = \gb^\dagger_s,\\
~ < 0, &\text{ if } \gb > \gb^\dagger_s.
\end{cases}
\end{align*}
\end{lem*}

\begin{figure*}
\centering
\begin{tikzpicture}
\begin{axis}[
scale=1.2,
width=12cm, height=8cm,
xlabel={$\gb$},
x label style={at={(axis description cs:1.05,0.9)}},
xmin=0, xmax=1.1,
ymin=-0.15, ymax=1,
axis lines=middle,
xtick=\empty,
ytick=\empty,
]
\addplot[domain=0:1, samples=100, black, thick] {max(0, min(1, (-0.025 + x * (0.0166667 + 4.44167 * x)) / (3 - x)^2))} node[right, pos=0.6] {$H_{\mu, F}(\gb|h,1)$};

\addplot[domain=0:1, samples=100, black, thick] {max(0, min(1, (-2.025 + x * (1.35 + 9.775 * x)) / (3 - x)^2))} node[right, pos=0.78] {$H_{\mu, F}(\gb|l,1)$};

\addplot[domain=0:1, samples=100, gray, dashed, thick] {max(0, min(1, (-0.975 + (13.9833 - 8.99722 * x) * x) / (3 - x)^2))} node[left, pos=0.6] {$H_{\mu, F}(\gb|h,0)$};

\addplot[domain=0:1, samples=100, gray, dashed, thick] {max(0, min(1, (-11.475 + (37.65 - 21.275 * x) * x) / (3 - x)^2))} node[left, pos=0.4] {$H_{\mu, F}(\gb|l,0)$};

\draw[dotted] (0.6923,0) node[below] {$\gb^\dagger_0$} -- (0.6923,0.825);

\draw[dotted] (0.5,0) node[below] {$\gb^\dagger_1$} -- (0.5,0.175);

\draw[solid] (0.073,-0.025) node[below] {$\ubar \gb^h$} -- (0.073,0.025);

\draw[solid] (0.3913,-0.025) node[below] {$\ubar \gb^l$} -- (0.3913,0.025);

\draw[dotted] (0.966,0) -- (0.966,1);

\draw[dotted] (0.7778,0) -- (0.7778,1);

\draw[solid] (0.966,-0.025) node[below] {$\bar \gb^h$} -- (0.966,0.025);

\draw[solid] (0.7778,-0.025) node[below] {$\bar \gb^l$} -- (0.7778,0.025);

\draw[dotted] (1,0) node[below] {$1$} -- (1,1);

\draw[dotted] (0,0) node[below right] {$0$} -- (0,0);

\node[right] at (0,0.95) {parameters:};
\node[right] at (0,0.88) {$\mu=0.6$, $p = 0.1$,};
\node[right] at (0,0.81) {$x(h)=0.9$, $x(l) = 0.4
$,};
\node[right] at (0,0.73) {$k(h)=k(l)=0$.};

\end{axis}
\end{tikzpicture}
\caption{\label{fig:Hmle} Distributions of state beliefs in \cref{eg:mle}}
\end{figure*}
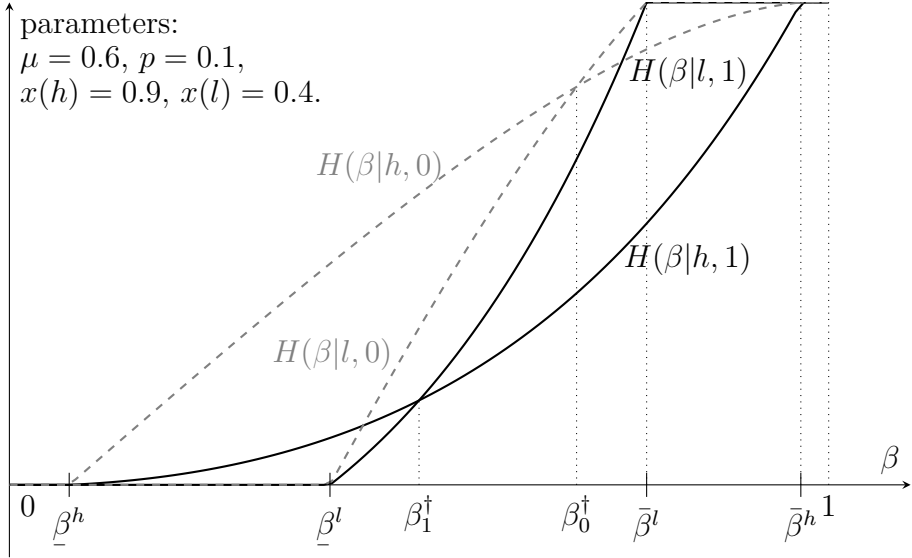

Thus, in each state $s$, the high type is more likely than the low type to have state belief short of $\gb$ if $\gb$ is less than $\gb^\dagger_s$, but the low type is more likely than the high type to have state belief short of $\gb$ if $\gb$ is above $\gb^\dagger_s$. This is because the low type is more likely than the high type to form an intermediate state belief. \cref{fig:Hmle} illustrates this.

\section{Informative equilibrium}
\label{sec:binaryc}

In this section, I characterize all informative equilibria and show that they need not be influential.

\subsection{Characterization}
\label{sec:characterization}

I first characterize the structure, existence, and essential uniqueness of informative equilibria. The following notations are essential. In each state $s$, I refer to $\gb^\dagger_s$ identified in \cref{lem:signals} as a crossing belief. Because this crossing belief depends on $(\mu, F)$, but not $p$, I sometimes write $\gb^\dagger_s$ as $\gb^\dagger_s(\mu,F)$. Let $\mathcal F$ be the set of experiment profiles satisfying $\gb^\dagger_1(\mu', F) < \gb^\dagger_0(\mu', F)$ for some $\mu' \in [\frac 1 2, 1)$, in addition to \cref{asp:lr2}. This inequality can be easily verified from the model primitives. Let $\Xi$ be the set of information structures $\xi=(\mu,p,F)$ in which $F \in \mathcal F$. 

The sets $\mathcal F$ and $\Xi$ are nonempty and nondegenerate. This is because for any $\mu \in [\frac 1 2, 1)$, $p \in (0,1)$, and $F=(F^h, F^l)$ satisfying \cref{asp:lr2}(a) and (b), if $F^h$ is sufficiently close to $F^*$, then \cref{asp:lr2}(c) holds and $\gb^\dagger_1 < \gb^\dagger_0$. The latter follows because $H_{\mu,F}(\cdot|h,0)$ is then sufficiently close to a distribution assigning a unit mass to belief zero so that $\gb^\dagger_0$ is sufficiently close to $\bar \gb^l$, $H_{\mu,F}(\cdot|h,1)$ is then sufficiently close to a distribution assigning a unit mass to belief one so that $\gb^\dagger_1$ is sufficiently close to $\ubar \gb^l$, and $\ubar \gb^l < \bar \gb^l$.

\begin{prop*}\label{prop:existence}
The following hold.
\begin{enumerate}\itemsep0em
\item In any informative equilibrium, there exists a unique $\gb \in (\gb^\dagger_1, \gb^\dagger_0)$ such that each sender type with state belief $b$ reports 1 if $b > \gb$ and 0 if $b < \gb$. Such $\gb$ satisfies
\begin{align}\label{eq:fixedpointbeta}
\gb r^*(0,1,\gb) + (1-\gb) r^*(0,0,\gb) = \gb r^*(1,1,\gb) + (1-\gb) r^*(1,0,\gb),
\end{align}
where for each state $s$,
\begin{align} 
\label{eq:r0gw}
r^*(0,s,\gb) &:=
\frac{p  H_{\mu,F}(\gb|h,s) 
}
{
\bE^p[H_{\mu,F}(\gb|t,s)]
},\\
r^*(1,s,\gb) &:= 
\frac{
p 
( 1- H_{\mu,F}(\gb|h,s) 
)
}
{
\bE^p[ 1- H_{\mu,F}(\gb|t,s) ]
},\label{eq:r1gw}
\end{align}
and the expectation $\bE^p[\cdot]$ is taken over $t$ with respect to the prior type distribution. The receiver matches his action with the sender's report if
\begin{align}\label{eq:advicequality}
\bE^p[ \mu (1-H_{\mu,F}(\gb|t,1)) + (1-\mu) H_{\mu,F}(\gb|t,0)] > \mu,
\end{align}
and takes action 1 irrespective of the sender's report otherwise. 
\item An informative equilibrium exists if and only if $F \in \mathcal F$. This equilibrium is essentially unique: all informative equilibria are outcome-equivalent.
\end{enumerate}
\end{prop*}

Part 1 states that in any informative equilibrium, each sender type reports 1 if her state belief exceeds a cutoff and reports 0 otherwise. Because in each state a correct report yields a strictly higher reputation than an incorrect one, and because each sender type views state 1 as more likely when her state belief $b$ is higher, there exists a cutoff $\gb$ such that she reports 0 if $b<\gb$ and 1 if $b>\gb$. This cutoff satisfies \eqref{eq:fixedpointbeta}, which requires the sender to be indifferent between the two reports at $b=\gb$ when the market correctly conjectures this cutoff strategy. This cutoff lies in $(\gb^\dagger_1,\gb^\dagger_0)$ because, by \cref{lem:signals}, the high type must be strictly more likely than the low type to have a belief below (resp., above) $\gb$ in state 0 (resp., state 1) for a correct report to yield a strictly higher reputation than an incorrect one in each state. The receiver who conjectures that the sender plays the cutoff strategy with threshold $\gb$ matches his action with report 1, as this report reinforces the prior state belief. In contrast, he need not match his action with report 0; he does so, and therefore the equilibrium is influential, if and only if \eqref{eq:advicequality} holds, which states that his payoff from matching his action with any report of the sender, or his matching payoff for short, strictly exceeds his reservation payoff.

Part 2 shows that $F \in \mathcal F$ is both necessary and sufficient for an informative equilibrium to exist and shows that if it exists, it is essentially unique. If $F \notin \mathcal F$, then $\gb^\dagger_1 \ge \gb^\dagger_0$ and so the sender's cutoff strategy characterizing an informative equilibrium does not exist. The proof of the converse proceeds as follows. I first show that if $F \in \mathcal F$, then $\gb^\dagger_1(\mu, F) < \gb^\dagger_0(\mu, F)$ for every $\mu$. This is because the value of the prior state belief does not affect the relative informativeness between the two types' experiments. Therefore, for any $\xi=(\mu,p,F)$ in which $F \in \mathcal F$, $\gb^\dagger_1 < \gb^\dagger_0$. I then show that a unique solution $\gb$ to \eqref{eq:fixedpointbeta} exists and lies in $(\gb^\dagger_1,\gb^\dagger_0)$; uniqueness of $\gb$ then implies that there is an essentially unique informative equilibrium. Intuitively, if $\gb$ is too high, then at state belief $\gb$, each sender type's expected reputation gain from reporting $1$ rather than $0$ is too large to keep her indifferent between the two reports: she views report $1$ as very likely to be correct, and the market conjecturing the sender's cutoff strategy  views a correct report $1$ as very likely to come from the sender type whose experiment yields sufficiently extreme signals leading to report 1, namely the high type. Similarly, $\gb$ cannot be too low, for otherwise each sender type's expected reputation gain from reporting $0$ rather than $1$ is too large to keep her indifferent between the two reports.



When no ambiguity arises, I denote by $\gb$ the (essentially unique) informative equilibrium in which the sender plays the above strategy with cutoff $\gb$ and refer to the corresponding strategy profile as a $\gb$-cutoff strategy profile. By \eqref{eq:fixedpointbeta}, $\gb$ depends on the information structure $\xi$; when useful, I write $\gb_\xi$ to emphasize this dependence. In general, the cutoff $\gb$ is different from $\frac 1 2$ and so the sender's report may be misleading to the receiver. Specifically, I say that the contrarian report may be misleading if $\gb > \frac 1 2$; in this case, the sender reports 0 even when her state belief lies in $(\frac 1 2, \gb)$ so that she considers state 1 more likely. Similarly, I say that the conformist report may be misleading if $\gb < \frac 1 2$; in this case, the sender reports 1 even when her state belief lies in $(\gb, \frac 1 2)$ so that she considers state 0 more likely. Note that in any informative equilibrium, only one of the two reports, but not both, may be misleading.


I next characterize how the prior state belief shapes the structure of the informative equilibrium, especially which report may be misleading. The following inequality is essential for the characterization; I interpret it momentarily.
\begin{align}\label{eq:fixedpointbetanew00}
\begin{multlined}[15cm]
\frac{1-H_{\frac 1 2, F}( \frac 1 2 |h,1)}
{
\bE^p[1-H_{\frac 1 2, F}( \frac 1 2 |t,1)]
} 
-
\frac{H_{\frac 1 2, F}( \frac 1 2 |h,1)}
{
\bE^p[H_{\frac 1 2, F}( \frac 1 2 |t,1)]
} 
\ge
\frac{H_{\frac 1 2, F}( \frac 1 2 |h,0)}
{
\bE^p[H_{\frac 1 2, F}( \frac 1 2 |t,0)]
} 
-
\frac{1-H_{\frac 1 2, F}( \frac 1 2 |h,0)}
{
\bE^p[1-H_{\frac 1 2, F}( \frac 1 2 |t,0)]
}.
\end{multlined}
\end{align}

\begin{prop*}\label{prop:cutoff}
Let $\gb_{\mu,p,F}$ be the informative equilibrium.
\begin{enumerate}\itemsep0em
\item $\gb_{\mu,p,F}$ is strictly increasing in $\mu$.
\item There is $\mu^I_{p,F} < 1/(1+\ubar \ell^h)$ such that $\gb_{\mu,p,F} > \frac 1 2$ if and only if $\mu > \mu^I_{p,F}$. Moreover,
\begin{align*}
\mu^I_{p,F} ~\begin{cases}
~> \frac 1 2, \quad &\text{ if \eqref{eq:fixedpointbetanew00} holds strictly},\\
~= \frac 1 2, \quad &\text{ if \eqref{eq:fixedpointbetanew00} binds},\\
~< \frac 1 2, \quad &\text{ if \eqref{eq:fixedpointbetanew00} fails}.
\end{cases}
\end{align*}
\end{enumerate}
\end{prop*}
Part 1 follows since a higher $\mu$ makes a correct report 0 a stronger signal of being the high type, as the high type is more likely than the low type to have state beliefs below $\gb_{\mu,p,F}$. This strengthens each type's incentive to report 0 for any state belief.

Part 2 shows that the contrarian report may be misleading if the prior state belief is sufficiently high and the conformist report may be misleading otherwise. Intuitively, $\mu^I_{p,F}$ is the prior state belief at which an informative equilibrium $\gb =\frac 1 2$ exists. In this equilibrium, each sender type reports truthfully the state she deems more likely. The left side of \eqref{eq:fixedpointbetanew00} is the sender's expected reputation gain from reporting $1$ in this equilibrium, whereas the right side is the report-$0$ counterpart. In this equilibrium, \eqref{eq:fixedpointbetanew00} must bind so that each sender type is indifferent between the two reports at state belief $\frac 1 2$. If \eqref{eq:fixedpointbetanew00} holds strictly, the incentive to report $1$ is too strong; maintaining this indifference requires the prior state belief to favor state 1 more than state 0 so that a correct report $0$ becomes a stronger reputational signal, implying $\mu^I_{p,F}>\frac 1 2$. The cases where \eqref{eq:fixedpointbetanew00} binds or fails are analogous. Consequently, for each $\mu>\mu^I_{p,F}$, part 1 implies that $\gb_{\mu,p,F} > \frac 1 2$, causing the contrarian report to be possibly misleading. The case for $\mu<\mu^I_{p,F}$ is analogous, resulting in  $\gb_{\mu,p,F} < \frac 1 2$ and a possibly misleading conformist report. Finally, to understand the strict upper bound $1/(1+\ubar \ell^h)$ on $\mu^I_{p,F}$, note that if $\mu^I_{p,F} \ge 1/(1+\ubar \ell^h)$, then no informative equilibrium $\gb=\frac 1 2$ exists at prior state belief $\mu^I_{p,F}$. If it did, then at this prior state belief, both types' state beliefs upon receiving their signals are at least $\frac 1 2$ by direct calculations. Because $\gb = \frac 1 2$, \cref{prop:existence} implies that in the equilibrium both types report 1 with probability one. The receiver's state belief is then unchanged upon receiving report 1, contradicting that this equilibrium is informative.

Part 2 implies that if \eqref{eq:fixedpointbetanew00} binds or fails and so $\mu^I_{p, F} \le \frac 1 2$, then in the informative equilibrium only the contrarian report may be misleading for all $\mu$; this includes the setting in which each sender type's experiment is symmetric across states, in which case $\mu^I_{p,F}=\frac 1 2$. On the other hand, because $\mu^I_{p,F} < 1/(1+\ubar \ell^h)$, \cref{asp:lr2}(c) implies that for sufficiently high $\mu$, only the contrarian report, but not the conformist report, may be misleading.

\cref{prop:cutoff} differs from the results in \citet{ottaviani2006professional,ottaviani2006reputational} that study the sender's optimal deviation incentives when the market expects her to report her signal truthfully. There are two key differences. First, in \cref{prop:cutoff}, the cutoff $\mu^I_{p, F}$ is determined by the sender's incentives when the market expects her to report the state she deems more likely, but not her signal. Second, \cref{prop:cutoff} characterizes how these incentives depend on the prior state belief.

\cref{prop:existence} and \cref{prop:cutoff} also generalize existing results. \citet[Proposition 7]{ottaviani2006reputational} show that an informative equilibrium exists where the sender plays a cutoff strategy and may send a misleading contrarian report under multiplicative linear experiments. \cref{prop:existence} shows that this existence relies on the fact that informative structures with multiplicative linear experiments lie in $\Xi$, and that the informative equilibrium is essentially unique. \cref{prop:cutoff} shows that contrarianism arises from symmetry of the multiplicative linear experiment across the two states.

\subsection{Influentialness}
\label{sec:noninfluential}

Finally, I show that the informative equilibrium need not be influential. By \cref{prop:existence}, an informative equilibrium is influential if and only if \eqref{eq:advicequality} holds, showing that influentialness centers on the cutoff $\gb_\xi$. Let $\Xi^*$ be the set of information structures $\xi$ in $\Xi$ for which $\mu > \mu^I_{p,F}$ and therefore $\gb_\xi > \frac 1 2$ so that the contrarian report may be misleading. By \cref{prop:cutoff}, $\Xi^*$ is nonempty and nondegenerate, containing for instance a neighborhood around information structures in which experiments are symmetric across states for both types.


\begin{prop*}[Influentialness]\label{prop:noninflusource}
The following hold.
\begin{enumerate}\itemsep0em
\item Let $\xi \in \Xi$. If $\xi \notin \Xi^*$, then the informative equilibrium is influential.
\item For any $F^h$, there exists an open set of $(p,\mu,F^l)$ on which $(p,\mu,(F^h,F^l)) \in \Xi^*$ and the informative equilibrium is noninfluential.
\end{enumerate}
\end{prop*}

Part 1 holds because in any informative equilibrium, as discussed, the receiver's best reply given report 1 is to take action 1. If $\xi \notin \Xi^*$, report 0 is not misleading by \cref{prop:cutoff} and so his best reply given report 0 is to take action 0.

Part 2 implies that even if the high type's experiment is arbitrarily close to perfect, the informative equilibrium can be noninfluential for informative structures in $\Xi^*$. For intuition, consider, for instance, an information structure in which the low type's experiment $F^l$ is sufficiently noisy that her state belief never favors state 0,\footnote{An example is the multiplicative experiment in \cref{eg:mle}, with low enough weight $x(l)$.} and the prior reputation $p$ is sufficiently low. The resulting informative equilibrium is noninfluential because the receiver perceives report 0 as too likely to be misleading. 

This noninfluentialness does not rely on the restriction to binary communication. It however relies on the receiver's action being discrete. If the receiver chose an action from a continuum and his payoff increased continuously as the action approached the true state, then the receiver would benefit in the informative equilibrium from even an infinitesimal amount of information transmitted by the sender; the equilibrium would then be influential.





\section{Complementarity}
\label{sec:complementarity}

In this section, I present my first main result, establishing a complementarity between high-quality prior information and the influentialness of communication. Recall that $\diff(\cdot,\cdot)$ denotes the Prokhorov metric, and that $F^*$ denotes the perfect experiment. 

\begin{thm*}[Complementarity]\label{thm:complementarity}
For each $p \in (0,1)$ and $(F^h, F^l) \in \mathcal F$, there exists $\mu^*_{p, F^l} \in [\frac 1 2, 1)$ such that for each $\mu \ge \mu^*_{p, F^l}$, there exists $\gd^*_{\mu, p, F^l}>0$ such that the informative equilibrium is influential if $\max_{s \in S} \diff( F(\cdot|h,s), F^*(\cdot|s) ) < \gd^*_{\mu, p, F^l}$.
\end{thm*}

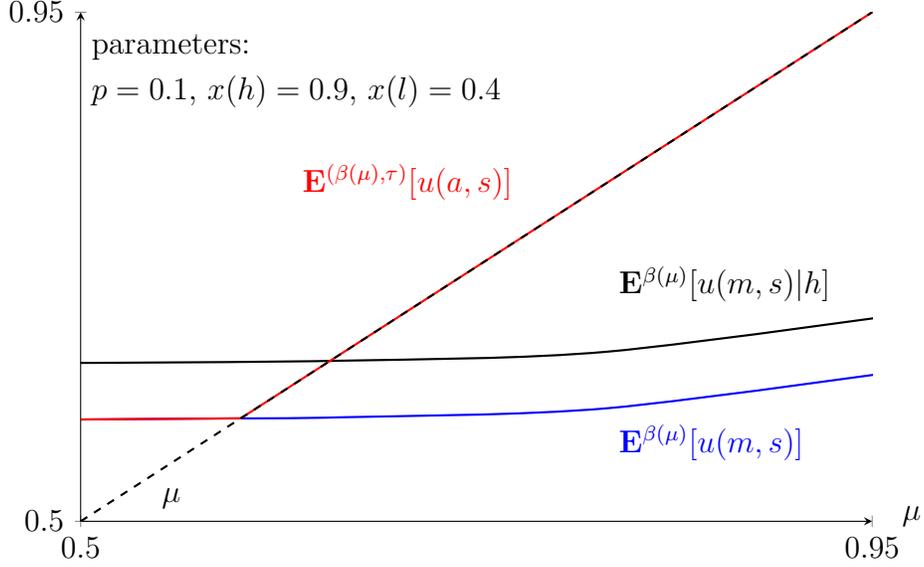
\begin{figure}
\centering
\begin{tikzpicture}
\begin{axis}[scale=1.2,
axis lines=left,
x label style={at={(axis description cs:1.05, 0.05)}},
xlabel={$\mu$},
ylabel={},
xmin=0.5, xmax=0.95,
ymin=0.5, ymax=0.95,
xtick={0.5,0.95},
xticklabels={0.5,$1/(1+\ubar \ell^h) = 1$},
ytick={0.5,0.95},
yticklabels={0.5,1},
width=10cm, height=7cm
]


\addplot[smooth, ultra thick] coordinates {
(0.5,0.54)  (0.565,0.565)
};
\addplot[smooth, ultra thick] coordinates {
(0.565,0.565) (0.65,0.62) (0.8, 0.8)
};
\addplot[smooth, ultra thick] coordinates {
(0.8, 0.8) (0.9,0.91) (1.0,0.971)
};

\node[right] at (0.685,0.64) {$\bE^{\gb_{\mu, p, F}}[u(m,s)]$};
\node[right] at (0.85,0.83) {$\mu$};
\node[right] at (0.5,0.93) {parameters:};
\node[right] at (0.5,0.89) {$p=0.1, x(l)=0.1, k=1000$.};

\addplot[dashed, thick] coordinates {
(0.5,0.5) (1.0,1.0)
};

\draw[dotted] (0.53,0.5) node [above right] {$\mu''$}-- (0.53,0.55) ;

\draw[dotted] (0.65, 0.5) node [above right] {$\mu'$} -- (0.65,0.65);

\draw[dotted] (0.9,0.5) node [above right] {$\mu'''$}-- (0.9,0.91);
\end{axis}
\end{tikzpicture}
\caption{Matching payoff versus reservation payoff in \cref{eg:hyper}}
\label{fig:illusfinal}
\end{figure}

\cref{thm:complementarity} shows that when the prior state belief is sufficiently high and the high type's experiment is sufficiently informative relative to it, the informative equilibrium is influential. \cref{fig:illusfinal} revisits \cref{eg:hyper} to illustrate this, plotting the receiver's
matching payoff and his reservation payoff in the informative equilibrium
against values of $\mu$ satisfying \cref{asp:lr2}(c), noting that the equilibrium is influential if and only if the former payoff strictly exceeds the latter. \cref{thm:complementarity} holds not because high prior state beliefs preclude report 0 from being misleading; \cref{prop:cutoff} has shown otherwise.

For intuition, consider a sequence of experiment profiles $(F^k)_{k=0}^\infty$, along which the low type's experiment is constant and the high type's experiment converges weakly to $F^*$. For each $\mu$ and sufficiently large $k$ relative to $\mu$, the information structure $(\mu, p, F^k)$ lies in $\Xi$, as explained in \cref{sec:characterization}. By \cref{prop:existence}, an (essentially unique) informative equilibrium $\gb_{\mu, p, F^k}$ exists; it is influential if and only if
\begin{align}\label{eq:formalize}
\bE^{\gb_{\mu,p,F^k}}[u(m,s)|h] &> \mu +  \frac{1-p}{p} \!\left( \mu - \bE^{\gb_{\mu,p,F^k}}[u(m,s)|l] \right),
\end{align}
namely that the receiver's matching payoff conditional on a high type strictly exceeds his reservation payoff $\mu$ plus some markup. This markup is proportional to the receiver's loss from matching his action with the low type's report relative to his reservation payoff. If the prior state belief $\mu$ is sufficiently high and in addition if, relative to $\mu$, the high type's experiment is sufficiently informative, \ie $k$ is sufficiently high, then the low type perceives that the high type is sufficiently likely to receive a signal strongly favoring state 1 and then, by \cref{prop:existence}, to report 1; in turn, to appear as a high type, the low type rarely reports 0. Thus, conditional on a low type, the receiver can virtually secure his reservation payoff by matching his action with the report. The markup on the right side of \eqref{eq:formalize} is then negligible, so that this right side is close to $\mu$ and lower than 1. Because the high type is sufficiently likely to receive signals inducing state beliefs strongly favoring the true state and then, by \cref{prop:existence}, to report the state she deems more likely to be true, the left side of \eqref{eq:formalize} is sufficiently close to one, and \eqref{eq:formalize} holds. \cref{thm:complementarity} then follows.

In \cref{thm:complementarity}, it is crucial that the prior state belief is sufficiently high. Part 2 of \cref{prop:noninflusource} has shown that even if the high type's experiment is arbitrarily close to perfect, there could be prior state beliefs that are not sufficiently high for which the informative equilibrium is noninfluential.

It is also crucial that the high type's experiment is sufficiently informative relative to the prior state belief. This is because sufficiently high prior state beliefs relative to the experiments render the informative equilibrium noninfluential:

\begin{prop*}\label{prop:highnoninflu}
Let $(\mu, p, F) \in \Xi$. There is $\bar \mu_{p,F} \in (\frac 1 2, 1/(1+\ubar \ell^h))$ such that if $\mu \ge \bar \mu_{p,F}$, then the informative equilibrium is noninfluential.
\end{prop*}

\cref{fig:illusfinal} also illustrates \cref{prop:highnoninflu}. In the figure, the experiment profile is fixed, and the informative equilibrium is noninfluential for sufficiently high prior state beliefs near the upper bound $1/(1+\ubar \ell^h)$ that follows from \cref{asp:lr2}(c). Intuitively, if the prior state belief attains this upper bound, the receiver is indifferent between matching his action with the sender's report and taking his ex ante optimal action, even if the sender is known to be a high type who reports truthfully. Hence, for nearby prior state beliefs, the receiver strictly prefers taking the ex ante optimal action irrespective of the sender's report because the sender is potentially a low type whose experiment is noisier than the high type's and may send a misleading contrarian report. Of course, by \cref{thm:complementarity}, the range $[\bar \mu_{p,F},1)$ of noninfluentialness collapses in the limiting case where the high type's experiment is perfect.

\cref{thm:complementarity} contrasts with the conventional wisdom that the receiver substitutes the sender's report for noisy prior information. This is also present in my model:

\begin{prop*}\label{prop:substitute}
Let $p \in (0,1)$ and $F \in \mathcal F$. There is $\ubar \mu_{p,F} \in (\frac 1 2, 1/(1+\ubar \ell^h))$ such that if $\mu < \ubar \mu_{p,F}$, then the informative equilibrium is influential.	
\end{prop*}

\cref{fig:illusfinal} also illustrates \cref{prop:substitute}. Intuitively, when prior information is sufficiently noisy, the receiver’s reservation payoff is low, so he optimally matches his action to the sender’s report, even though the report, whether contrarian or conformist, may be misleading. 




Taken together, both sufficiently noisy prior information and sufficiently precise prior information help sustain influentialness, but they do so via different mechanisms. \cref{prop:compare} clarifies that the substitution mechanism and the complementarity mechanism have fundamentally different payoff implications:

\begin{prop*}\label{prop:compare}
There exists a set of information structures $\tilde \Xi^* \subseteq \Xi$, with nonempty interior, such that for each $\xi' = (\mu', p, F) \in \tilde \Xi^*$, there exist $\xi'' = (\mu'', p, F), \xi''' = (\mu''', p, F) \in \tilde \Xi^*$ with $\mu'' < \mu' < \mu'''$ such that
\begin{enumerate}\itemsep0em
\item The informative equilibrium is influential under $\xi''$ and $\xi'''$, but not under $\xi'$.
\item $\bE^{\gb_{\xi''}}[u(a,s)] < \bE^{\gb_{\xi'}}[u(a,s)] < \bE^{\gb_{\xi'''}}[u(a,s)]$.
\end{enumerate}
\end{prop*}

\cref{fig:illusfinal} also illustrates \cref{prop:compare}. Part 1 follows from \cref{prop:noninflusource}, \cref{thm:complementarity}, and \cref{prop:substitute}. Part 2 holds because the receiver fares better when relying on prior information than on the sender if, in the latter setting, prior information is sufficiently noisy that he relies on the sender's report despite the risk of being misled, reflecting the substitution mechanism; he fares worse if, in the latter setting, prior information is sufficiently informative that the contrarian report is rarely misleading, reflecting the complementarity mechanism.

In general, solving for the largest set of prior state beliefs over which the informative equilibrium is influential, for arbitrary initial reputations and experiment profiles, is intractable. Doing so requires solving for all roots of the difference between the receiver’s informative-equilibrium matching payoff and his reservation payoff, $\bE^{\gb_{\mu,p,F}}[u(m,s)] - \mu$, as a function of $\mu$ on $[\frac{1}{2},1)$. In \cref{fig:illusfinal}, this function has exactly three roots, yielding the four-part partition of the prior state belief space: the informative equilibrium is influential in the lowest and second-highest regions, but not in the other two.

\section{Allocative efficiency}
\label{sec:cs}

In any informative equilibrium $\gb$, the receiver's payoff $\bE^\gb[u(a,s)]$ measures allocative efficiency and is equal to the maximum between his matching payoff $\bE^\gb[u(m,s)]$ and his reservation payoff $\mu$. In this section, I present my second main result, characterizing how this payoff depends on the quality of prior information.

\begin{thm*}[Allocative efficiency and prior information]\label{thm:highermuhigherpayoff}
Let $(F^k)_{k=0}^\infty$ be a sequence of experiment profiles in $\mathcal F$ along which the low type's experiment is constant and the high type's experiment converges weakly to $F^*$. Let $\mu^I_p := \lim_{k\to\infty}\mu^I_{p,F^k}$, where $\mu^I_{p,F^k}$ is identified in \cref{prop:cutoff}. Then:
\begin{enumerate}\itemsep0em
\item For each $\mu \in [\frac1 2,\mu^I_p)$, there exists $K_\mu$ such that for all $k\ge K_\mu$, the receiver's informative-equilibrium payoff is strictly decreasing in $\mu$.
\item For each $\mu \in (\mu^I_p,1)$, there exists $K_\mu$ such that for all $k\ge K_\mu$, the receiver's informative-equilibrium payoff is strictly increasing in $\mu$.
\end{enumerate}
\end{thm*}

\cref{thm:highermuhigherpayoff} shows that when the high type's experiment is sufficiently informative, the receiver's informative-equilibrium payoff is essentially single-dipped in the prior state belief. The experiment profile in \cref{eg:hyper} satisfies the theorem's conditions.

\cref{thm:highermuhigherpayoff} yields sharp policy implications. If prior information is sufficiently noisy that the conformist report may be misleading, marginal improvements in it harm the receiver. Otherwise, these improvements unambiguously benefit the receiver---these improvements may or may not render the informative equilibrium noninfluential, as \cref{prop:compare} suggests. If it does, then the receiver's gain comes from his reliance on the better prior information; otherwise, his gain comes from his reliance on the sender's report that outperforms the better prior information. \cref{thm:highermuhigherpayoff} further suggests that improvements in prior information should be sufficiently aggressive to avoid the perverse outcome in which they reduce the receiver’s payoff. This is because the receiver's informative-equilibrium payoff is strictly increasing over a range of sufficiently high prior state beliefs and tends to his first-best payoff of one as the prior state belief approaches one; this convergence follows because the receiver's informative-equilibrium payoff is at least his reservation payoff, which tends to one.

In the special case in which each sender type's experiment is symmetric across states along the sequence $(F^k)_{k=0}^\infty$, by \cref{prop:cutoff}, the range of prior state beliefs in part 1 is empty because then $\mu^I_{p, F^k}=\frac 1 2$ for each $k$. Part 2 then implies that the receiver's informative-equilibrium payoff is strictly increasing over all prior state beliefs when the high type's experiment is sufficiently informative.

For intuition of \cref{thm:highermuhigherpayoff}, consider any sequence of experiments $(F^k)_{k=0}^\infty$ as stated in the theorem. It is instructive to first consider, for each $k$, the derivative of the receiver's informative-equilibrium matching payoff $\bE^{\gb_{\mu,p,F^k}}[u(m,s)]$ conditional on each sender type $t$ with respect to the prior state belief $\mu$, given by
\begin{align}\label{eq:typechange}
\begin{multlined}[t][13cm]
\overbrace{1 - F^k(\ell_{\mu,p,F^k}|t,1)}^{\textnormal{term A}}
-
\overbrace{F^k(\ell_{\mu,p,F^k}|t,0)}^{\textnormal{term B}}
\\[0.5em]
+
\underbrace{\frac{\partial \ell_{\mu,p,F^k}}{\partial \mu}}_{\textnormal{term C}} \underbrace{\vphantom{\frac{\partial \ell_{\mu,p,F^k}}{\partial \mu}}\left((1-\mu)f^k(\ell_{\mu,p,F^k}|t,0)-\mu f^k(\ell_{\mu,p,F^k}|t,1)\right)}_{\textnormal{term D}}.
\end{multlined}
\end{align}
where $\ell_{\mu,p,F^k}$ is the likelihood ratio corresponding to the equilibrium cutoff $\gb_{\mu,p,F^k}$, namely $\gb_{\mu,p,F^k}= \mu \ell_{\mu,p,F^k}/(\mu \ell_{\mu,p,F^k}+1-\mu)$. Expression \eqref{eq:typechange} captures two channels through which a marginal increase in $\mu$ changes the receiver's matching payoff conditional on each type. First, this payoff then places greater weight on matching with a correct report $1$ and less on matching with a correct report $0$, captured by term A minus term B. Second, $\ell_{\mu,p,F^k}$ changes, altering the probability that each report is issued, captured by the product of terms C and D.

Several observations concerning \eqref{eq:typechange} lead to \cref{thm:highermuhigherpayoff}. First, term C is negative. Intuitively, more precise prior information strengthens the reputation gain associated with a correct report 0; sustaining this stronger reputation gain in equilibrium requires that a correct report 0 is less likely to be sent by the low type, or equivalently a lower $\ell_{\mu,p,F^k}$ corresponding to a lower state belief cutoff $\gb_{\mu,p,F^k}$.

Second, for sufficiently large $k$, the derivative \eqref{eq:typechange} conditional on the high type is negligible. Intuitively, conditional on the high type, because the experiment is nearly perfect, by \cref{prop:existence}, both the probability of a correct report $1$ and that of a correct report $0$ are close to $1$. Therefore the difference between terms A and B is negligible. Next, by denoting local $L^1$ convergence of a sequence of functions $(w^k)_{k=0}^\infty$ to function $w$ on a domain $D$ by ``$w^k \ra w$ in $L^1_{\mathrm{loc}}(D)$,'' weak convergence of the high type's experiment to $F^*$ implies that the high type's signal densities satisfying, as $k \to \infty$, $f^k(\cdot|h,0) \to 0 \text{ in } L^1_{\textnormal{loc}}((0, \infty])$ and $f^k(\cdot|h,1) \to 0 \text{ in } L^1_{\textnormal{loc}}([0, \infty))$.\footnote{This is an implication of the Portmanteau theorem (\egg \citealp[Theorem 2.1]{billingsley2013convergence}). Of course, this implication holds here because the limiting distribution $F^*(\cdot|s)$ is a singular distribution for each state $s$; in general, weak convergence of a distribution need not imply local $L^1$ convergence of its density.} Intuitively, in the weak limit, all probability mass must concentrate arbitrarily close to $0$ in state $0$ and to $\infty$ in state $1$, requiring the density to vanish everywhere else. Therefore, for sufficiently high $k$, term D, and therefore the product of terms C and D, is also negligible, as $\gb_{\mu,p,F^k} \in (\gb^\dagger_1, \gb^\dagger_0)$ and therefore $\ell_{\mu,p,F^k}>0$ so that her experiment places vanishing mass near the cutoff $\ell_{\mu,p,F^k}$.

Third, for any $\mu \in [\frac 1 2, \mu^I_p)$ and sufficiently large $k$, the derivative \eqref{eq:typechange} conditional on the low type is negative. In this case, $\mu \in [\frac 1 2, \mu^I_{p,F^k})$, so report 1 may be misleading. This is sustained in equilibrium because a correct report 1 carries a stronger reputational signal than a correct report 0 does. Given that the market perceives the high type as almost always reporting correctly, a correct report 1 must therefore be less likely than a correct report 0 to be sent by the low type. Hence, conditional on the low type, term A minus term B in \eqref{eq:typechange} is negative; in addition, the lower chance of the low type sending report 0 given a higher $\mu$ by the first observation above implies that the product of terms C and D is also negative.\footnote{Formally, because report 1 may be misleading, the state belief cutoff is $\gb_{\mu,p,F^k} < \frac 1 2$. This implies that in \eqref{eq:typechange}, term D is negative, and so is the product of terms C and D.} 

Fourth, for any $\mu \in (\mu^I_p, 1)$ and sufficiently large $k$, the derivative \eqref{eq:typechange} conditional on the low type is positive. Here, report 0 may be misleading. Consequently,  for reasons analogous to the third observation above, conditional on the low type, both term A minus term B and the product of terms C and D are positive.


Taken together, for any $\mu \in [\frac 1 2, \mu^I_p)$ and sufficiently large $k$, a marginal increase in $\mu$ strictly decreases the receiver's matching payoff. This then implies that the marginal increase in $\mu$ strictly decreases his informative-equilibrium payoff, since this payoff coincides with the matching payoff when the equilibrium is influential, which holds on $[\frac 1 2, \mu^I_{p,F^k})$ by part 1 of \cref{prop:noninflusource}. Consequently, part 1 of \cref{thm:complementarity} follows. Analogously, for any $\mu \in (\mu^I_p,1)$ and sufficiently large $k$, a marginal increase in $\mu$ strictly increases the receiver's matching payoff. Since his informative-equilibrium payoff is the maximum of his matching payoff and reservation payoff $\mu$, it also strictly increases with $\mu$. This leads to part 2 of \cref{thm:highermuhigherpayoff}.

\section{Reputation and prior information}
\label{sec:initrep}

My previous results have shown how prior information shapes influentialness and allocative efficiency. In this section, I show how the sender's initial reputation $p$ interacts with prior information in determining both. For any $p \in (0,1)$ and $F \in \mathcal F$, let $I_{p,F} \subseteq [\frac 1 2, 1/(1+\ubar \ell^h) )$ denote the set of prior state beliefs satisfying \cref{asp:lr2}(c) over which the informative equilibrium is influential.

\begin{prop*}\label{prop:alloeffrep}
Let $(F^k)_{k=0}^\infty$ be a sequence of experiment profiles in $\mathcal F$ along which the low type's experiment is constant and the high type's experiment converges weakly to $F^*$. There exists $K > 0$ such that for each $k \ge K$,  
\begin{enumerate}\itemsep0em
\item An increase in the sender's initial reputation strictly expands the region of prior state beliefs over which the informative equilibrium is influential: for any $p, p' \in (0,1)$, $p < p'$ implies $I_{p,F^k} \subsetneq I_{p',F^k}$.
\item A marginal increase in $p$ weakly increases the receiver's informative-equilibrium payoff, and strictly so if and only if the resulting equilibrium is influential.
\end{enumerate}
\end{prop*}

Combining \cref{prop:alloeffrep} with \cref{thm:complementarity} and \cref{thm:highermuhigherpayoff} yields further policy implications for allocative efficiency. When the conformist report may be misleading, reputation improvements dominate prior-information improvements because the latter, unless sufficiently aggressive, harm the receiver. Otherwise, reputation improvements weakly benefit the receiver and are generally less effective than prior-information improvements, which strictly benefit the receiver; nonetheless, reputation improvements complement the latter by expanding the range of prior state beliefs that sustain influentialness and raising the receiver's payoff within that range.


Part 1 follows from two observations. First, \cref{prop:substitute} implies that for any experiment profile $F$, sufficiently high prior state beliefs render the informative equilibrium noninfluential, so that $I_{p,F}$ is a strict subset of $[\frac 1 2, 1/(1+\ubar \ell^h))$. Second, for sufficiently large $k$ and experiment profile $F^k$, an increase in the sender's initial reputation strictly raises the receiver's informative-equilibrium matching payoff, expanding the range of prior state beliefs $I_{p, F^k}$ over which this matching payoff strictly exceeds the receiver's reservation payoff. To see this, for any function $\phi(t)$, let $\Delta \phi := \phi(h)-\phi(l)$. The derivative of the matching payoff with respect to $p$ is
\begin{align}\label{eq:derivativepmain}
\hspace*{-25px}\begin{multlined}[13cm]
\overbrace{\Delta\!\left[\mu (1-F^k(\ell_{\mu,p,F^k}|t,1)) + (1-\mu) F^k(\ell_{\mu,p,F^k}|t,0) \right]}^{\textnormal{term A}} \\[0.5em]
+ \underbrace{\frac{\partial \ell_{\mu,p,F^k}}{\partial p} \bE^p\!\left[(1-\mu) f^k(\ell_{\mu,p,F^k}|t,0)-\mu f^k(\ell_{\mu,p,F^k}|t,1)\right]}_{\textnormal{term B}},
\end{multlined}
\end{align}
where, as in \cref{sec:cs}, $\ell_{\mu,p,F^k}$ is the likelihood ratio corresponding to the equilibrium cutoff $\gb_{\mu,p,F^k}$. This expression is positive, leading to part 1, because a marginal increase in $p$ raises the matching payoff through two channels. First, it increases the weight on the high type's report relative to the low type's, captured by term A. This raises the matching payoff since the high type's experiment Blackwell-dominates the low type's. Second, the likelihood ratio $\ell_{\mu,p,F^k}$ adjusts, changing the distribution of reports, captured by term B. As the high type's experiment is nearly perfect, the market views her report as almost always correct; incorrect misleading reports therefore result in reputations close to zero. An increase in $p$ reduces each sender type's marginal reputation gain from a correct report, thereby causing the cutoff $\ell_{\mu,p,F^k}$ to shift in a way that reduces the probability of the misleading report being issued. If $\mu=\mu^I_{p,F^k}$, neither report is misleading, so term B is zero; otherwise, either the conformist report or the contrarian report may be misleading, and term B is positive.


Finally, part 2 follows because that the receiver's informative-equilibrium payoff equals his matching payoff when the equilibrium is influential, which is strictly increasing in the sender's initial reputation, and equals his reservation payoff otherwise, which is independent of the initial reputation.

\section{Concluding comments}
\label{sec:extensions}

In this paper, I have characterized the allocative implications of prior information in a canonical setting of reputational cheap talk. I conclude by discussing how my results extend to several variants of the model relevant to different applications.

\paragraph{Private communication.} In my main model, the sender's message is observable to the market, as in canonical setups. This assumption may be inappropriate when, for example, the sender is a private corporate analyst. Accordingly, consider a variant in which the sender's message is hidden from the market and the market infers the sender's type only from the receiver's action; the model is otherwise identical.

In this alternative model, my main results continue to hold with suitable modifications. In any influential equilibrium, the receiver's action matches the sender's report, so the market perfectly infers the report. Thus, when an influential equilibrium exists, it has the same structure as in the main model and is essentially unique; the conditions for its existence are also as in the main model. \cref{thm:complementarity} therefore extends. However, when an influential equilibrium does not exist, the sender's informative-equilibrium strategy is generally undetermined because her reports do not affect her reputation. Nonetheless, in this case, the receiver's payoff equals the reservation payoff, as in the main model. Consequently, \cref{thm:highermuhigherpayoff} extends, with the receiver's informative-equilibrium payoff replaced by his highest attainable equilibrium payoff.


Notably, in this alternative model, focusing on binary messages in characterizing influential equilibria entails no loss of generality since the sender's message affects her payoff only through the receiver's binary action; see \cref{sec:binaryprivate} for details.

\paragraph{Delegation.} In some applications, the decision maker chooses whether to delegate his action to an expert rather than acting himself. For example, a defendant chooses whether to settle or let a lawyer take his case to trial. Consider a variant of my model in which the receiver chooses whether to delegate decision-making to the sender.\footnote{To minimize departure from the main model, I refer to the decision maker as the receiver and the expert as the sender, although this alternative model is no longer a communication game.} If he does not delegate, the receiver takes an action himself as in the main model. If he delegates, the sender acts on the receiver's behalf, and the market evaluates her value using this action, the eventual revelation of the state, and its conjecture about each sender type's strategy. Here, each sender type's incentive to choose action $a=s$ upon delegation is identical to her incentive to report state $s$ in the influential equilibrium of the main model. Consequently, she chooses action 0 if her state belief falls below $\gb$ and action 1 otherwise, where $\gb$ is characterized as in \cref{prop:existence}. In any such equilibrium, the receiver delegates if \eqref{eq:advicequality} holds and does not delegate otherwise. The equilibrium structure is effectively identical to that in the private-communication variant above, and my results extend similarly.

\paragraph{Choosing whether to obtain advice.} My model has followed the canonical setup in assuming that the receiver must receive a message from the sender. My results extend to a version of my model in which the receiver can choose whether to obtain a report at the outset, for reasons analogous to the delegation variant above.



\paragraph{Symmetric type uncertainty.} Following \cite{ottaviani2006professional}, consider a variant of my model where the sender, like the receiver, initially assigns probability $p$ to her type being high and $1-p$ to her type being low. Let $L(h)=L(l)=L\equiv[\ubar\ell,\bar\ell]$, so the sender cannot perfectly infer her type from her signal. Her strategy is a measurable function $\gs:L\to\Delta(S)$ mapping signals to report distributions.

In this alternative model, my main results extend following suitable modifications. In any informative equilibrium, part 1 of \cref{prop:existence} characterizing the sender's and the receiver's strategies hold because the sender's payoff is determined as in the main model. Thus, provided that an informative equilibrium exists, it is essentially unique and has the same structure characterized as in the main model. However, part 2 of \cref{prop:existence} no longer holds. Because the sender forms her state belief without knowing whether her signal is drawn from $F^h$ or $F^l$, the solution $\gb$ to \eqref{eq:fixedpointbeta}, which characterizes the sender's belief cutoff in the informative equilibrium, may lie below her lowest attainable state belief, ruling out the existence of an informative equilibrium. This non-existence result generalizes \citet[Proposition 8]{ottaviani2006professional} in the setting of multiplicative linear experiments. Consequently, the essential single-dippedness in \cref{thm:highermuhigherpayoff} applies only to the range of prior state beliefs bounded away from one over which an informative equilibrium exists. On the other hand, for each $\mu$, if the average sender's experiment is sufficiently close to perfect relative to $\mu$, an essentially unique informative equilibrium exists and the remaining results, including the complementarity in \cref{thm:complementarity}, extend.

\pagebreak
\appendix
\appendixpage
\appendixtitletocoff

\section{Omitted details}
\label{sec:od}

\subsection{\cref{eg:mle}}
\label{sec:multiplicativelinearappendix}


In this Appendix, I prove that multiplicative linear experiments satisfy \cref{asp:lr2}(a) and (b). For each type $t$, the support of signals is
\begin{align*}
L(t)=\left[\frac{1-x(t)}{1+x(t)},\frac{1+x(t)}{1-x(t)}\right].
\end{align*}
Because $x(l)<x(h)$, \cref{asp:lr2}(a) holds.

Next, by direct calculations, for each type $t$ and state $s$,
\begin{align*}
F(\ell| t,s)
=
\begin{cases} 
(1+x(t)) \dfrac{(1+x(t))\ell-(1-x(t))}{2x(t)(1+\ell)} - x(t) \!\left(\dfrac{(1+x(t))\ell-(1-x(t))}{2x(t)(1+\ell)} \right)^2, &\quad \text{ if } s =0,\\
(1-x(t))\dfrac{(1+x(t))\ell-(1-x(t))}{2x(t)(1+\ell)} + x(t) \!\left(\dfrac{(1+x(t))\ell-(1-x(t))}{2x(t)(1+\ell)} \right)^2, &\quad \text{ if } s =1.
\end{cases}
\end{align*}
Consequently,
\begin{align*}
\frac{f(\ell| t,1)}{F(\ell| t,1)}
&=
\frac{8\ell}
{(1+\ell)(3\ell+1-x(t)(1+\ell))(x(t)(1+\ell)+\ell-1)},\\
\frac{f(\ell| t,0)}{F(\ell| t,0)}
&=
\frac{8}
{(1+\ell)(x(t)(1+\ell)+\ell-1)(x(t)(1+\ell)+\ell+3)},\\
\frac{f(\ell| t,1)}{1-F(\ell| t,1)}
&=
\frac{8\ell}
{(1+\ell)(x(t)(1+\ell)-(\ell-1))(x(t)(1+\ell)+(3\ell+1))},\\
\frac{f(\ell| t,0)}{1-F(\ell| t,0)}
&=
\frac{8}
{(1+\ell)((3+\ell)-x(t)(1+\ell))(x(t)(1+\ell)-(\ell-1))}.
\end{align*}
It is then straightforward to verify that each expression is strictly decreasing in $x(t)$ on $(0,1]$. Therefore \cref{asp:lr2}(b) holds.

\subsection{\cref{eg:hyper}}
\label{sec:hyperexponentialappendix}

In this Appendix, I prove the claims in \cref{eg:hyper}. 

I first show that for sufficiently large $k$, \cref{asp:lr2} holds. Part (a) follows because $[\ubar \ell^h, \bar \ell^h] = [0,\infty]$ whereas $[\ubar\ell^l, \bar \ell^l] = [(1-x(l))/(1+x(l)), (1+x(l))/(1-x(l))]$, with $0 < x(l) <1$. Part (c) also follows because the upper bound $1/(1+\ubar \ell^h)=1$. To prove part (b), by direct computation, writing $b_k := k/(k^2-k+1)$, for every $\ell>0$,
\begin{align*}
\frac{f^k(\ell|h,0)}{1-F^k(\ell|h,0)}
=
\frac{
\left(1-\frac1k\right)k \expo^{-k\ell}
+
\frac{1}{k} b_k \expo^{-b_k \ell}
}{
\left(1-\frac1k\right) \expo^{-k\ell}
+
\frac1k \expo^{-b_k \ell}
} &\le 
\frac{\left(1-\frac1k\right)k \expo^{-k\ell}
+
\frac{1}{k} b_k \expo^{-b_k \ell}}{\frac1k \expo^{-b_k \ell}} \\
&= (k^2-k) \expo^{- k \ell + b_k \ell} + b_k \ra 0, \quad \text{ as } k \ra \infty.
\end{align*}
Similarly,
\begin{align*}
\frac{f^k(\ell|h,1)}{1-F^k(\ell|h,1)} &\le \frac{(k^2-k) \ell \expo^{-k \ell + b_k \ell}}{\ell + 1/b_k} + b_k\frac{\ell}{\ell + 1/b_k} \ra 0, \quad \text{ as } k \ra \infty,\\
\frac{f^k(\ell|h,0)}{F^k(\ell|h,0)} &\le 2 \!\left[ 
(k-1) \expo^{-k \ell} + \frac{b_k}{k} \expo^{-b_k \ell}
\right]\! \ra 0, \quad \text{ as } k \ra \infty,\\
\frac{f^k(\ell|h,1)}{F^k(\ell|h,1)} &\le 3 \ell (k^2-k) \expo^{-k \ell} + 3 \ell b_k \expo^{-b_k \ell} \ra 0, \quad \text{ as } k \ra \infty.
\end{align*}
Because $\ubar \ell^l>0$ and the hazard rate and the reverse hazard rate of the low type are both bounded away from zero in each state, as derived in \cref{sec:multiplicativelinearappendix}, it follows that part (b) holds for sufficiently high $k$.

Next, to prove that $F^k(\cdot|h,s)$ converges weakly to $F^*(\cdot|s)$ for each state $s$, by the Portmanteau theorem (\egg \citealp[Theorem 2.1]{billingsley2013convergence}), it suffices to prove that $f^k(\cdot|h,0) \to 0 \text{ in } L^1_{\textnormal{loc}}((0, \infty])$ and $f^k(\cdot|h,1) \to 0 \text{ in } L^1_{\textnormal{loc}}([0, \infty))$ as $k \ra \infty$.\footnote{Recall from \cref{sec:cs} that I denote local $L^1$ convergence of a sequence of functions $(w^k)_{k=0}^\infty$ to function $w$ on a domain $D$ by ``$w^k \ra w$ in $L^1_{\mathrm{loc}}(D)$.''
} Let $0<c<c'$. Then
\begin{align*}
\int_c^{c'} f_k(\ell|h,0)\,d\ell
&=
\left(1-\frac{1}{k}\right)\int_c^{c'} k \expo^{-k\ell} \diff \ell
+
\frac{1}{k(k-1+\frac{1}{k})}\int_c^{c'} \expo^{-(k-1+1/k)\ell} \diff \ell \\
&\le \left(1-\frac{1}{k}\right) \expo^{- k c} + \frac{c'-c}{k(k-1+\frac{1}{k})} \ra 0,\quad \text{ as } k \ra \infty.
\end{align*}
Hence $f^k(\cdot|h,0) \ra 0$ in $L^1_{\textnormal{loc}}((0, \infty])$. Similarly, 
\begin{align*}
\int_c^{c'} f_k(\ell|h,1) \diff \ell
&\le
\int_0^{c'} \ell f_k(\ell|h,0) \diff \ell \\
&= \left(1-\frac{1}{k}\right)\int_c^{c'} k \ell \expo^{-k\ell} \diff \ell
+
\frac{1}{k(k-1+\frac{1}{k})}\int_c^{c'} \ell \expo^{-(k-1+1/k)\ell} \diff \ell \\
&\le \left(1-\frac{1}{k}\right) \frac 1 k + \frac{(c')^2}{2k(k-1+\frac 1 k)} \ra 0, \quad \text{ as } k \ra \infty.
\end{align*}
Hence $f^k(\cdot|h,1) \ra 0$ in $L^1_{\textnormal{loc}}([0, \infty))$, as desired.

\section{Proofs}
\label{sec:proofs}

\subsection{Proof of claim in \cref{fn:symmetry}}
\label{sec:fnsymmetry}

As in \cref{fn:g} in \cref{eg:mle}, denote the density of $G$ by $g$ and the likelihood ratio by $\gl(y|t)=g(y|t,1)/g(y|t,0)$. Suppose that the monotone likelihood ratio property holds, namely $\gl(\cdot|t)$ is strictly increasing, and so the inverse $\gl^{-1}(\cdot|t)$ is well-defined. Symmetry requires $1-G(1-y|t,1)=G(y|t,0)$ for each $y \in [0,1]$. Differentiating yields $g(y|t,0)=g(1-y|t,1)$. Because this holds for every $y\in[0,1]$, it follows that $g(1-y|t,0)=g(y|t,1)$. Then, the likelihood ratio satisfies
\begin{align*}
\gl(1-y|t) = \frac{g(1-y|t,1)}{g(1-y|t,0)} = \frac{g(y|t,0)}{g(y|t,1)} = \frac{1}{\gl(y|t)}.
\end{align*}
This implies $\gl^{-1}(1/\ell|t) = 1-\gl^{-1}(\ell|t)$, and so
\begin{align*}
F(1/\ell|t,1) = G( \gl^{-1}(1/\ell|t)|t,1 ) = G( 1-\gl^{-1}(\ell|t)|t,1) = 1 - G( \gl^{-1}(\ell|t)|t,0) = 1 - F(\ell|t,0),
\end{align*}
as desired.

\subsection{Proof of \cref{lem:infthenrep}}

Fix any informative equilibrium $(\gs, \tau)$, and consider each sender type's best reply to the market's conjecture $\gs$. Given state belief $b$, the sender's payoff from reporting 0 is
\begin{align}\label{eq:0payoff}
b r(0,1,\gs) + (1-b) r(0,0,\gs),
\end{align}
her payoff from reporting 1 is
\begin{align}\label{eq:1payoff}
b r(1,1,\gs) + (1-b) r(1,0,\gs).
\end{align}
She prefers report 1 to 0 if 
\begin{align}\label{eq:fixedpointbeta0}
b [ r(1,1,\gs) - r(0,1,\gs)] \ge (1 - b) [ r(0,0,\gs) - r(1,0,\gs)]. 
\end{align}
and prefers report 0 to 1 if.
\begin{align}\label{eq:fixedpointbeta00}
b [ r(1,1,\gs) - r(0,1,\gs)] \le (1 - b) [ r(0,0,\gs) - r(1,0,\gs)].
\end{align}
With probability one, $b \in (0,1)$. I show that if \eqref{eq:correctreport} fails, then \eqref{eq:correctreport2} holds. Suppose then that \eqref{eq:correctreport} fails and $r(1,1, \gs) - r(0,1,\gs) \le 0$. I consider five cases in order.
\begin{enumerate}\itemsep0em
\item If $r(1,1, \gs) - r(0,1,\gs) \le 0$ and $r(0,0, \gs) - r(1,0,\gs) > 0$, then for each $b \in (0,1)$, the left side of \eqref{eq:fixedpointbeta0} is nonpositive and the right side is positive, and so the sender's best reply is to report 0. Because the market's conjecture of each sender type's strategy must be equal to this type's best reply in equilibrium, by part 1 of \cref{def:eqm}, the strategy profile $\gs$ must be that each type reports 0 with probability one irrespective of her signal, so that the receiver's state belief is $\mu \ge \frac 1 2$ upon receiving this report and has no strict incentive to take action 0; a contradiction.
\item If $r(1,1,\gs) - r(0,1,\gs) < 0$ and $r(0, 0,\gs) - r(1,0,\gs) = 0$, then as in Case 1, the equilibrium strategy $\gs$ must specify that each sender type reports 0 given each state belief $b \in (0,1)$, again yielding a contradiction.
\item If $r(1,1,\gs) - r(0,1,\gs) = 0$ and $r(0, 0,\gs) - r(1,0,\gs) < 0$, then the equilibrium strategy $\gs$ must specify that each sender type reports 1 given each state belief $b \in (0,1)$,  again yielding a contradiction.
\item If $r(1,1,\gs) - r(0,1,\gs) = 0$ and $r(0,0,\gs) - r(1,0,\gs) = 0$, then by writing $\bP^\gs[m|t,s]$ as the probability that report $m$ is sent conditional on type $t$ and state $s$ in the equilibrium, $r(0,0,\gs) = r(1,0,\gs)$, or equivalently,
\begin{align*}
\frac{p \bP^\gs[0|h,0]}{p \bP^\gs[0|h,0] + (1-p) \bP^\gs[0|l,0]}
&=
\frac{p \bP^\gs[1|h,0]}{p \bP^\gs[1|h,0] + (1-p) \bP^\gs[1|l,0]}.
\end{align*}
This implies 
\begin{align}\label{eq:P0hl}
\bP^\gs[0|h,0] = \bP^\gs[0|l,0]  
\end{align}
as $\bP^\gs[1|t,0] = 1-\bP^\gs[0|t,0]$ for each type $t$. Similarly, $r(1,1,\gs) = r(0,1,\gs)$, which implies
\begin{align}\label{eq:P1hl}
\bP^\gs[1|h,1] = \bP^\gs[1|l,1].
\end{align}
Consequently, \eqref{eq:P0hl} and \eqref{eq:P1hl} imply
\begin{align*}
r(0,0,\gs) = r(1,0,\gs) = r(1,1,\gs) = r(0,1,\gs) = p,
\end{align*}
which further implies $\bP^\gs[1|h,1] = \bP^\gs[1|h,0]$ and $\bP^\gs[0|h,1] = \bP^\gs[0|h,0]$.
In this equilibrium, the receiver's belief that the state is 1 upon receiving report 0 is therefore
\begin{align*}
\frac{\mu ( p \bP^\gs[0|h,1] + (1-p) \bP^\gs[0|l,1] )}
{\bE^\mu[ p \bP^\gs[0|h,s] + (1-p) \bP^\gs[0|l,s] ]
}
&= \frac{\mu \bP^\gs[0|h,1]}
{\mu \bP^\gs[0|h,1] +
(1-\mu) \bP^\gs[0|h,0] 
} \\
&= \mu \\
&\ge \frac 1 2,
\end{align*}
contradicting that the receiver has a strict incentive to match his action with report 0 in the equilibrium.
\item The remaining case $r(1,1,\gs) - r(0,1,\gs) < 0$ and $r(0,0,\gs) - r(1,0,\gs) < 0$ is \eqref{eq:correctreport2}. Therefore, if \eqref{eq:correctreport} fails, then \eqref{eq:correctreport2} holds.
\end{enumerate}
The case where \eqref{eq:correctreport} fails and $r(0,0, \gs) - r(1,0,\gs) \le 0$ is analogous. Finally, by analogous arguments, if \eqref{eq:correctreport2} fails, then \eqref{eq:correctreport} holds. Therefore in the equilibrium, either \eqref{eq:correctreport} or \eqref{eq:correctreport2} holds.

\subsection{Proof of \cref{lem:signals}}

Recall \eqref{eq:HF}. For each $(t,s)$, $H_{\mu, F}(\gb|t,s) = 0$ if $\gb \le \ubar \gb^t$ and $H_{\mu, F}(\gb | t,s) =1$ if $\gb \ge \bar \gb^t$. Moreover, because $0 \le \ubar \gb^h < \ubar \gb^l < \bar \gb^l < \bar \gb^h \le 1$, for each $s$, $H_{\mu,F}(\gb|h,s) - H_{\mu,F}(\gb|l,s) > 0$ for each $\gb \in (\ubar \gb^h, \ubar \gb^l]$ and $H_{\mu,F}(\gb|h,s) - H_{\mu,F}(\gb|l,s) < 0$ for each $\gb \in [\bar \gb^l, \bar \gb^h)$. Therefore, to complete the proof, it suffices to show that $H_{\mu, F}(\gb|h,s) - H_{\mu, F}(\gb|l,s)$ is strictly decreasing in $\gb$ on $(\ubar \gb^l, \bar \gb^l)$. Indeed, for any $\gb, \gb' \in (\ubar \gb^l, \bar \gb^l)$ with $\gb' > \gb$,
\begin{align*}
&~ ( H_{\mu, F}(\gb'|h,s) - H_{\mu, F}(\gb'|l,s)) - ( H_{\mu, F}(\gb|h,s) - H_{\mu, F}(\gb|l,s)) \\
=&~ \int_{\frac{1-\mu}{\mu} \frac{\gb}{1-\gb}}
^
{
\frac{1-\mu}{\mu} \frac{\gb'}{1-\gb'}
}
f(\ell|h,s) - f(\ell|l,s)
\diff \ell < 0,
\end{align*}
as was to be shown, where the inequality follows because for each $\ell \in (\ubar \ell^l, \bar \ell^l)$ and state $s$, \eqref{eq:hrates} implies that $f(\ell|l,s) > f(\ell|h,s)$.

\subsection{Proof of \cref{prop:existence}}

The following lemma is essential.

\begin{lem*}\label{lem:mlrpimplication}
Fix $\mu$ and $F$. For each $t$ and $\gb \in (\ubar \gb^t, \bar \gb^t)$, $H_{\mu, F}(\gb|t,0) > H_{\mu, F}(\gb|t,1)$.
\end{lem*}
\begin{proof}[Proof of \cref{lem:mlrpimplication}] 
Fix type $t$. For each $\ell,\ell' \in (\ubar \ell^t, \bar \ell^t)$ satisfying $\ell' > \ell$. Rewriting this inequality with the identities $\ell = f(\ell|t,1)/f(\ell|t,0)$ and $\ell' = f(\ell'|t,1)/f(\ell'|t,0)$ yields
\begin{align}\label{eq:mlrp0}
f(\ell'|t,1) f(\ell|t,0) > f(\ell'|t,0) f(\ell|t,1). 
\end{align}
Integrating both sides of \eqref{eq:mlrp0} over $\ell$ from $0$ to $\ell'$, and rearranging, yields
\begin{align*}
\ell' = \frac{f(\ell'|t,1)}{f(\ell'|t,0)} > \frac{F(\ell'|t,1)}{F(\ell'|t,0)}.
\end{align*}
Similarly, integrating both sides of \eqref{eq:mlrp0} over $\ell'$ from $\ell$ to $1$, and rearranging, yields
\begin{align*}
\frac{1 - F(\ell|t,1)}{1- F(\ell|t,0)} > \frac{f(\ell|t,1)}{f(\ell|t,0)} = \ell.
\end{align*}
Because $\ell$ and $\ell'$ are arbitrarily chosen, it follows that for every $\ell \in (\ubar \ell^t, \bar \ell^t)$,
\begin{align*}
\frac{1 - F(\ell|t,1)}{1- F(\ell|t,0)}
> \ell > 
\frac{F(\ell|t,1)}{F(\ell|t,0)},
\end{align*}
implying $F(\ell|t,0) > F(\ell|t,1)$. Therefore, by \eqref{eq:HF}, for any $\gb \in (\ubar \gb^t, \bar \gb^t)$,
\begin{align*}
H_{\mu, F}(\gb|t,0) = 
F\!\left( 
\frac{1-\mu}{\mu} \frac{\gb}{1-\gb}
\middle| t,0 
\right) > F\!\left( 
\frac{1-\mu}{\mu} \frac{\gb}{1-\gb}
\middle| t,1 
\right) = 
H_{\mu, F}(\gb|t,1),
\end{align*}
as was to be shown.
\end{proof}

I prove the two parts of \cref{prop:existence} in order.

\begin{enumerate}\itemsep0em
\item Fix an informative equilibrium $(\gs, \tau)$, and consider each sender type's best reply to the market's conjecture $\gs$. Given state belief $b$, each sender type prefers reporting 1 to reporting 0 if  \eqref{eq:fixedpointbeta0} holds and prefers reporting 0 to reporting 1 if  \eqref{eq:fixedpointbeta00} holds. In addition, \eqref{eq:correctreport} holds and so the left side of \eqref{eq:fixedpointbeta0} is strictly increasing in $b$ while the right side of \eqref{eq:fixedpointbeta0} is strictly decreasing in $b$. Moreover, at $b=0$, the left side of \eqref{eq:fixedpointbeta0} is zero while the right side is positive, whereas at $b=1$, the left side of \eqref{eq:fixedpointbeta0} is positive while the right side is zero. Because both sides of \eqref{eq:fixedpointbeta0} are continuous in $b$, by the intermediate value theorem, there exists $\gb \in (0,1)$ such that the sender is indifferent between the two reports when her state belief is $\gb$, strictly prefers to report 0 if $b < \gb$ and 1 if $b > \gb$. Because the market's conjecture of each sender type's strategy must be equal to this sender type's best reply to the market's conjecture in equilibrium by part 1 of \cref{def:eqm}, the payoffs \eqref{eq:r0gw} and \eqref{eq:r1gw} follow and the indifference condition \eqref{eq:fixedpointbeta} must hold at state belief $\gb$. In addition, by \eqref{eq:correctreport},  $r^*(0,0,\gb) > r^*(1,0,\gb)$ and $r^*(1,1,\gb) > r^*(0,1,\gb)$, or equivalently, $H_{\mu,F}(\gb|h,0) > H_{\mu,F}(\gb|l,0)$ and $H_{\mu,F}(\gb|h,1) < H_{\mu,F}(\gb|l,1)$. These latter inequalities imply that $\gb < \gb^\dagger_0$ and $\gb > \gb^\dagger_1$ by \cref{lem:signals}. 

The receiver's best reply when conjecturing the sender's strategy profile with cutoff $\gb$ is to match his action with report 1. To see this, note that upon receiving report 1, the receiver forms state belief equal to
\begin{align}\label{eq:receiverinterim1}
\frac
{\mu  \bE^p[ 1 - H_{\mu, F}(\gb|t,1) ] }
{\mu \bE^p[ 1 - H_{\mu, F}(\gb|t,1) ] + (1-\mu) \bE^p[ 1 - H_{\mu, F}(\gb|t,0)]}.
\end{align}
Because $\gb \in (\gb^\dagger_1, \gb^\dagger_0)$ and so for each type $t$, $H_{\mu, F}(\gb|t,1) < 
H_{\mu, F}(\gb|t,0)$,  \eqref{eq:receiverinterim1} is higher than $\mu \ge \frac 1 2$. In turn, because \eqref{eq:advicequality} states that the receiver's payoff from matching his action with any report is strictly higher than his reservation payoff, the receiver's best reply is to match his action with the sender's report if \eqref{eq:advicequality} holds and is to take action 1 irrespective of the sender's report to obtain his reservation payoff otherwise.
\item Part 1 already shows that if an informative equilibrium exists, then $\gb^\dagger_1 < \gb^\dagger_0$ and so $F \in \mathcal F$. Consider then the converse direction. Suppose that $F \in \mathcal F$, and so $\gb^\dagger_1(\mu', F) < \gb^\dagger_0(\mu', F)$ for some $\mu' \in [\frac 1 2, 1)$. I first show that $\gb^\dagger_1(\mu, F) < \gb^\dagger_0(\mu, F)$ for each $\mu \in [\frac 1 2, 1)$. For each state $s$ and $\mu \in [\frac 1 2, 1)$, because $\gb^\dagger_s(\mu, F)$ is the unique $\gb \in (\ubar \gb^l, \bar \gb^l)$ solving $H_{\mu, F}(\gb|h,s) - H_{\mu, F}(\gb|l,s) = 0$ by \cref{lem:signals}, or equivalently,
\begin{align}\label{eq:FF}
F\!\left(\frac{1-\mu}{\mu} \frac{\gb}{1-\gb} \middle| h,s \right)\! - 
F\!\left(\frac{1-\mu}{\mu} \frac{\gb}{1-\gb} \middle| l,s \right)\! = 0,
\end{align}
an application of the implicit function theorem on \eqref{eq:FF} yields
\begin{align*}
\frac{\partial \gb^\dagger_s(\mu, F)}
{\partial 
\mu } &=
- 
\frac
{ -\frac{\gb^\dagger_s(\mu, F)}{1-\gb^\dagger_s(\mu, F)} \frac{1}{\mu^2} \!\left[ f\!\left(\frac{1-\mu}{\mu} \frac{\gb^\dagger_s(\mu, F)}{1-\gb^\dagger_s(\mu, F)} \middle| h,s \right)\! - 
f\!\left(\frac{1-\mu}{\mu} \frac{\gb^\dagger_s(\mu, F)}{1-\gb^\dagger_s(\mu, F)} \middle| l,s \right)\! \right]\! }
{
\frac{1-\mu}{\mu} \frac{1}{(1-\gb^\dagger_s(\mu, F))^2} \!\left[ f\!\left(\frac{1-\mu}{\mu} \frac{\gb^\dagger_s(\mu, F)}{1-\gb^\dagger_s(\mu, F)} \middle| h,s \right)\! - 
f\!\left(\frac{1-\mu}{\mu} \frac{\gb^\dagger_s(\mu, F)}{1-\gb^\dagger_s(\mu, F)} \middle| l,s \right)\! \right]\!
} \\[0.25em]
&= \frac{(1-\gb^\dagger_s(\mu, F)) \gb^\dagger_s(\mu, F)}{(1-\mu)\mu}.
\end{align*}
Solving this differential equation yields 
\begin{align*}
\gb^\dagger_s(\mu, F) = 
\frac{\gb^\dagger_s(\mu', F) \mu (1-\mu')}{\gb^\dagger_s(\mu', F) (\mu - \mu') + \mu' (1-\mu)},
\end{align*}
which, alongside the assumption $\gb^\dagger_1(\mu', F) < \gb^\dagger_0(\mu', F)$, implies $\gb^\dagger_0(\mu, F) - \gb^\dagger_1(\mu, F) > 0$ by direct calculation, as desired. In the following, fix some $(\mu, p, F) \in \Xi$. Note that by \eqref{eq:HF}, \eqref{eq:fixedpointbeta} can be rewritten as
\begin{align}\label{eq:fixedpointbetanew}
\begin{multlined}[12cm]
\gb \bigg[
\frac{p (1-F(\frac{1-\mu}{\mu}\frac{\gb}{1-\gb}|h,1))}{
\bE^p[1-F(\frac{1-\mu}{\mu}\frac{\gb}{1-\gb}|t,1)]} 
-
\frac{p F(\frac{1-\mu}{\mu}\frac{\gb}{1-\gb}|h,1)}{\bE^p[ F(\frac{1-\mu}{\mu}\frac{\gb}{1-\gb}|t,1)]}
\bigg] 
\\[0.25em]
=(1-\gb) \bigg[
\frac{p F(\frac{1-\mu}{\mu}\frac{\gb}{1-\gb}|h,0) }{\bE^p[F(\frac{1-\mu}{\mu}\frac{\gb}{1-\gb}|t,0)]
}
-
\frac{p (1-F(\frac{1-\mu}{\mu}\frac{\gb}{1-\gb}|h,0))}{\bE^p[1-F(\frac{1-\mu}{\mu}\frac{\gb}{1-\gb}|t,0)]} 
\bigg].
\end{multlined}
\end{align}
At $\gb = \gb^\dagger_1$, the left side of \eqref{eq:fixedpointbetanew} is zero whereas the right side is positive. At $\gb = \gb^\dagger_0$, the left side of this inequality is positive whereas the right side is zero. By continuity of both sides in $\gb$, there exists at least one $\gb \in (\gb^\dagger_1, \gb^\dagger_0)$ satisfying \eqref{eq:fixedpointbetanew}. Such $\gb$ is unique because the left side of \eqref{eq:fixedpointbetanew} is strictly increasing in $\gb$ and the right side of \eqref{eq:fixedpointbetanew} is strictly decreasing in $\gb$, which follows because, by direct calculation and \cref{asp:lr2}, for each state $s$,
\begin{align*}
&~ \frac{\partial}{\partial \gb} \!\left( \frac{F(\frac{1-\mu}{\mu}\frac{\gb}{1-\gb}|h,s)}{F(\frac{1-\mu}{\mu}\frac{\gb}{1-\gb}|l,s)} \right) \\
=&~ 
\frac{1-\mu}{(1-\gb)^2 \mu} \frac{
F(\frac{1-\mu}{\mu}\frac{\gb}{1-\gb}|l,s) f(\frac{1-\mu}{\mu}\frac{\gb}{1-\gb}|h,s)
-
F(\frac{1-\mu}{\mu}\frac{\gb}{1-\gb}|h,s)
f(\frac{1-\mu}{\mu}\frac{\gb}{1-\gb}|l,s)
}{F(\frac{1-\mu}{\mu}\frac{\gb}{1-\gb}|l,s)^2}\\
<&~ 0,
\end{align*}
and
\begin{align*}
&~ \frac{\partial}{\partial \gb} \!\left( \frac{1-F(\frac{1-\mu}{\mu}\frac{\gb}{1-\gb}|h,s)}{1-F(\frac{1-\mu}{\mu}\frac{\gb}{1-\gb}|l,s)} \right)\\
=&~ -
\frac{1-\mu}{(1-\gb)^2 \mu} \frac{
(1-F(\frac{1-\mu}{\mu}\frac{\gb}{1-\gb}|l,s)) f(\frac{1-\mu}{\mu}\frac{\gb}{1-\gb}|h,s)
-
(1-F(\frac{1-\mu}{\mu}\frac{\gb}{1-\gb}|h,s))
f(\frac{1-\mu}{\mu}\frac{\gb}{1-\gb}|l,s)
}{(1-F(\frac{1-\mu}{\mu}\frac{\gb}{1-\gb}|l,s))^2}\\
>&~ 0.
\end{align*}
Because the cutoff $\gb$ characterizing the sender's strategy profile in any informative equilibrium is uniquely identified, the receiver has a unique best reply, except in the non-generic case where $\gb$ is such that the receiver is indifferent between action 1 and action 0 upon receiving report 0, in which case his best reply upon this report is undetermined. Thus,  if an informative equilibrium exists, then it is essentially unique.
\end{enumerate}

\subsection{Proof of \cref{prop:cutoff}}

I prove the two parts in order.

\begin{enumerate}
\item Recall that \eqref{eq:fixedpointbeta} can be rewritten as \eqref{eq:fixedpointbetanew}. Analogous to the proof of \cref{prop:existence}, by direct calculations, holding $\gb$ fixed, the left side of \eqref{eq:fixedpointbetanew} is strictly decreasing in $\mu$, whereas the right side of \eqref{eq:fixedpointbetanew} is strictly increasing in $\mu$, because for each state $s$,
\begin{align*}
\frac{\partial}{\partial \mu} \!\left( \frac{F(\frac{1-\mu}{\mu}\frac{\gb}{1-\gb}|h,s)}{F(\frac{1-\mu}{\mu}\frac{\gb}{1-\gb}|l,s)} \right) &> 0 \quad \text{ and } \quad 
\frac{\partial}{\partial \mu} \!\left( 
\frac{1-F(\frac{1-\mu}{\mu}\frac{\gb}{1-\gb}|h,s)}
{1-F(\frac{1-\mu}{\mu}\frac{\gb}{1-\gb}|l,s)} \right) < 0.
\end{align*}
Holding $\mu$ fixed, the left side of \eqref{eq:fixedpointbetanew} is strictly increasing in $\gb$ and the right side of \eqref{eq:fixedpointbetanew} is strictly decreasing in $\gb$ as shown in the proof of \cref{prop:existence}. Therefore $\gb_{\mu, p, F}$ satisfying \eqref{eq:fixedpointbetanew} is strictly increasing in $\mu$.
\item I show that there exists a unique $\mu^I \equiv \mu^I_{p, F} < 1/(1+\ubar \ell^h)$, characterized by
\begin{align}\label{eq:cc_char}
\begin{multlined}[12cm]
\frac{1-H_{\mu^I, F}( \frac 1 2 |h,1)}
{
\bE^p[1-H_{\mu^I, F}( \frac 1 2 |t,1)]
} 
-
\frac{H_{\mu^I, F}( \frac 1 2 |h,1)}
{
\bE^p[H_{\mu^I, F}( \frac 1 2 |t,1)]
} 
= 
\frac{H_{\mu^I, F}( \frac 1 2 |h,0)}
{
\bE^p[H_{\mu^I, F}( \frac 1 2 |t,0)]
} 
-
\frac{1-H_{\mu^I, F}( \frac 1 2 |h,0)}
{
\bE^p[1-H_{\mu^I, F}( \frac 1 2 |t,0)]
},
\end{multlined}
\end{align}
such that $\gb_{\mu,p,F} < \frac 1 2$ if and only if $\mu < \mu^I$. Note first that at $\mu = \mu^I$ and $\gb = \frac 1 2$, \eqref{eq:fixedpointbetanew} simplifies to \eqref{eq:cc_char}. Next, note that $\gb_{\mu,p,F} \ra 1$ as $\mu \ra 1$, because $\gb_{\mu,p,F} > \ubar \gb^l(\mu,F^l)$ and because
\begin{align*}
\ubar \gb^l(\mu,F^l) = \frac{\mu f(\ubar \ell^l|l,1)}{\mu f(\ubar \ell^l|l,1) + (1-\mu) f(\ubar \ell^l|l,0)} \ra 1
\end{align*}
as $\mu \ra 1$. Consider the inequality \eqref{eq:fixedpointbetanew00}. The left (resp., right) side of \eqref{eq:fixedpointbetanew00} is equal to the left (resp., right) side of \eqref{eq:fixedpointbetanew} evaluated at $\mu=\gb=\frac 1 2$. Because the left (resp., right) side of \eqref{eq:fixedpointbetanew} is strictly decreasing (resp., increasing) in $\mu$ holding $\gb$ fixed, if \eqref{eq:fixedpointbetanew00} holds strictly, then $\mu^I > \frac 1 2$. If \eqref{eq:fixedpointbetanew00} binds, then $\mu^I = \frac 1 2$. Finally, if \eqref{eq:fixedpointbetanew00} fails to hold, then $\mu^I < \frac 1 2$. It remains to show that $\mu^I < 1/(1+\ubar \ell^h)$. Suppose towards a contradiction that $\mu^I \ge 1/(1+\ubar \ell^h)$. By definition of $\mu^I$, as explained in the main text, there is an informative equilibrium $\gb=\frac 1 2$ at $\mu=\mu^I$. By \cref{asp:lr2}(c), the high type's state belief is at least $\frac 1 2$ irrespective of her signal, and with probability one, strictly higher than $\frac 1 2$. Then, by \cref{asp:lr2}(a), the low type's state belief is strictly higher than $\frac 1 2$ irrespective of her signal. Consequently, in this equilibrium, report 1 is sent by each type with probability one. The receiver's state belief given report 1 then remains as $\mu$, contradicting that the equilibrium is informative. 
\end{enumerate}

\subsection{Proof of \cref{prop:noninflusource}}

I prove the two parts in order.
\begin{enumerate}
\item Fix $\xi \in \Xi$ and an informative equilibrium $\gb_\xi$.  Suppose $\xi \notin \Xi^*$ so that $\gb_\xi \le \frac 1 2$. Because $\gb_\xi \in (\gb^\dagger_1, \gb^\dagger_0)$, from the proof of part 1 of \cref{prop:existence}, the receiver's best reply upon receiving report 1 is to take action 1. I show that upon receiving report 0 instead, the receiver's best reply is to take action 0. To this end, it suffices to show that
\begin{align*}
\frac
{p H_{\mu,F}(\gb_\xi|h,0) + (1-p) H_{\mu,F}(\gb_\xi|l,0)}
{p H_{\mu,F}(\gb_\xi|h,1) + (1-p) H_{\mu,F}(\gb_\xi|l,1)} > \frac{\mu}{1-\mu},
\end{align*}
because then, \eqref{eq:advicequality} holds. Let
\begin{align*}
z := \frac{1-\mu}{\mu} \frac{\gb_\xi}{1-\gb_\xi}.
\end{align*}
Note that for each type $t \in \{h,l\}$,
\begin{align*}
\frac{F(z|t,1)}{F(z|t,0)} = \frac{\int_0^z f(\ell|t,1) \diff \ell}{\int_0^z f(\ell|t,0) \diff \ell} = 
\frac{\int_0^z \ell f(\ell|t,0) \diff \ell}{\int_0^z f(\ell|t,0) \diff \ell} = \bE[\ell|\ell \le z, t,s=0] < z,
\end{align*}
where the second equality uses the identity $\ell = f(\ell|t,1)/f(\ell|t,0)$. Because $\gb_\xi \le \frac 1 2$,
\begin{align*}
\frac{F(z|t,0)}{F(z|t,1)} > \frac{1}{z} = \frac{\mu}{1-\mu} \frac{1-\gb_\xi}{\gb_\xi} \ge \frac{\mu}{1-\mu} \quad \implies \quad F(z|t,0) > \frac{\mu}{1-\mu} F(z|t,1).
\end{align*}
As a result,
\begin{align*}
p F(z|h,0) + (1-p) F(z|l,0) > \frac{\mu}{1-\mu} ( p F(z|h,1) + (1-p) F(z|l,1)),
\end{align*}
implying
\begin{align*}
\frac
{p H_{\mu,F}(\gb_\xi|h,0) + (1-p) H_{\mu,F}(\gb_\xi|l,0)}
{p H_{\mu,F}(\gb_\xi|h,1) + (1-p) H_{\mu,F}(\gb_\xi|l,1)} = \frac{p F(z|h,0) + (1-p) F(z|l,0)}{p F(z|h,1) + (1-p) F(z|l,1)} > \frac{\mu}{1-\mu},
\end{align*}
as desired, where the equality uses \eqref{eq:HF}.
\item I proceed via a series of lemmas.

\begin{lem*}\label{lem:maxat12}
In any informative equilibrium $\gb$, for each type $t$, the receiver's matching payoff conditional on type $t$, namely
\begin{align}\label{eq:matchingpayofft}
\bE^\gb[u(m,s)|t] = (1-\mu) H_{\mu, F}(\gb|t,0) + \mu 
(1-H_{\mu, F}(\gb|t,1)),
\end{align}
evaluated at $\gb = \frac 1 2$ is strictly higher than evaluated at any $\gb \in (\frac 1 2, 1]$.
\end{lem*}

\begin{proof}[Proof of \cref{lem:maxat12}]
Let $u^*(t)$ denote the receiver's payoff if he knew that the sender's type is $t$ and observed her signal before optimally taking an action. Call this the receiver's first-best action. Note that for each type $t$, $\bE^{\gb }[u(m,s)|t] \le u^*(t)$ and $\bE^{\frac 1 2}[u(m,s)|t] = u^*(t)$. Conditional on each $t$, if $\gb > \frac 1 2$, then for each state belief $b > \gb$ or $b < \frac 1 2$, the receiver's action in the equilibrium would coincide with his first-best action. If $b \in (\frac 1 2, \gb)$, the receiver's action in the equilibrium differs from his first-best action. Because $\gb \in (\gb^\dagger_1, \gb^\dagger_0) \subsetneq (\ubar \gb^l, \bar \gb^l) \subsetneq (\ubar \gb^h, \bar \gb^h)$, each type's state belief lies in $(\frac 1 2, \gb)$ with positive probability. Consequently, $\bE^{\gb}[u(m,s)|t] < u^*(t) = \bE^{\frac 1 2}[u(m,s)|t]$ if $\gb > \frac 1 2$.
\end{proof}

\begin{lem*}\label{lem:nonemptynonvaluelow}
For each $\phi \in (0, \frac 1 4)$, there is an open set of experiments $\mathcal F_\phi$ in $\mathcal F$ such that for each $F=(F^h, F^l) \in \mathcal F_\phi$ and  $\mu \in [ \frac 1 2 + \phi, 1 - \phi]$, $\ubar \gb^l(\mu,F^l) > \frac 1 2$.
\end{lem*}

\begin{proof}[Proof of \cref{lem:nonemptynonvaluelow}]
For every $\phi \in (0, \frac 1 4)$, define $\mathcal F_\phi$ as the set of $F$ in $\mathcal F$ given which
\begin{align*}
\frac{(\frac 1 2 + \phi) f(\ubar \ell^l|l,1)}{(\frac 1 2 + \phi) f(\ubar \ell^l|l,1) + (\frac 1 2 - \phi) f(\ubar \ell^l|l,0)} > \frac 1 2  > \frac{(1- \phi) f(\ubar \ell^h|h,1)}{(1- \phi) f(\ubar \ell^h|h,1) + \phi f(\ubar \ell^h|h,0)}.
\end{align*}
The set $\mathcal F_\phi$ has a nonempty interior in $\mathcal F$: the above inequalities do not contradict parts (a) and (b) of \cref{asp:lr2}, and any $F$ satisfying these inequalities satisfies \cref{asp:lr2}(c) because
\begin{align*}
\ubar \gb^h(\mu,F^h) &= \frac{\mu f(\ubar \ell^h|h,1)}{\mu f(\ubar \ell^h|h,1) + (1-\mu) f(\ubar \ell^h|l,1)} \\
&\le \frac{(1 - \phi) f(\ubar \ell^h|h,1)}{(1-\phi) f(\ubar \ell^h|h,1) + \phi f(\ubar \ell^h|h,1)} < \frac 1 2,
\end{align*}
and so $\ubar \ell^h < (1-\mu)/\mu$. Given any $F \in \mathcal F_\phi$, for each $\mu \in [\frac 1 2 + \phi, 1-\phi]$, 
\begin{align*}
\ubar \gb^l(\mu,F^l) &= \frac{\mu f(\ubar \ell^l|l,1)}{\mu f(\ubar \ell^l|l,1) + (1-\mu) f(\ubar \ell^l|l,0)} \\
&\ge \frac{(\frac 1 2 + \phi) f(\ubar \ell^l|l,1)}{(\frac 1 2 + \phi) f(\ubar \ell^l|l,1) + (\frac 1 2 - \phi) f(\ubar \ell^l|l,0)} > \frac 1 2,
\end{align*}
where the equality follows by definition of $\ubar y^l$ and the last inequality uses the definition of $\mathcal F_\phi$, as desired.
\end{proof}

Hereafter, fix $\phi \in (0, \frac 1 4)$. Fix some $\mu^*_\phi \in (\frac 1 2 + \phi, 1-\phi)$.

\begin{lem*}\label{lem:lowp}
Let $F \in \mathcal F_\phi$. There exists $\bar p_{\mu^*_\phi, F}>0$ such that for each $p \in (0, \bar p_{\mu^*_\phi, F}]$, there is an open interval $B_{p,F}(\mu^*_\phi) \subsetneq [\frac 1 2, 1)$ containing $\mu^*_\phi$ such that for each $\mu \in B_{p,F}(\mu^*_\phi)$, \begin{align}\label{eq:noninflumu}
\bE^{\gb_{\mu,p,F}}[u(m,s)] < \mu.
\end{align}
\end{lem*}

\begin{proof}[Proof of \cref{lem:lowp}]
Let $F \in \mathcal F_\phi$. Consider a sequence $(p^n)_{n=0}^\infty$, where $p^n \in (0,1)$ for each $n$, converging monotonically to 0. I show that for sufficiently large $n$,
\begin{align}\label{eq:proofstrict}
\begin{multlined}[t][10cm]
\bE^{p^n} \big[ \mu^*_\phi (1-H_{\mu^*_\phi, F}(\gb_{\mu^*_\phi, p^n, F}|t,1)) + 
(1-\mu^*_\phi) H_{\mu^*_\phi, F}(\gb_{\mu^*_\phi, p^n, F}|t,0) \big] < \mu^*_\phi.
\end{multlined}
\end{align}
This is because in the limit as $n \ra \infty$, the left side of \eqref{eq:proofstrict} tends to
\begin{align*}
&~ \lim_{p \downarrow 0}  \mu^*_\phi (1-H_{\mu^*_\phi, F}( \gb_{\mu^*_\phi, p, F} | l,1 ) ) + (1-\mu^*_\phi) H_{\mu^*_\phi, F}( \gb_{\mu^*_\phi, p, F}  | l, 0 ) \\
<&~ \mu^*_\phi \!\left(1-H_{\mu^*_\phi, F} \!\left( \textstyle \frac 1 2 \middle| l,1 \right)\! \right)\! + (1-\mu^*_\phi) H_{\mu^*_\phi, F}\!\left( \textstyle \frac 1 2  \middle| l, 0 \right)\! = \mu^*_\phi,
\end{align*}
where the inequality follows because, conditional on $t=l$, \eqref{eq:matchingpayofft} evaluated at $\gb=\frac 1 2$ is strictly higher than \eqref{eq:matchingpayofft} evaluated at any $\gb > \frac 1 2$ by \cref{lem:maxat12} and because $\lim_{p \downarrow 0} \gb_{\mu^*_\phi, p, F} \ge \gb^\dagger_1(\mu^*_\phi, F) > \ubar \gb^l(\mu^*_\phi, F) > \frac 1 2$ by \cref{lem:signals}, \cref{prop:existence}, and \cref{lem:nonemptynonvaluelow}, and the equality follows because $\ubar \gb^l(\mu^*_\phi, F) > \frac 1 2$ so that $H_{\mu^*_\phi, F}( \frac 1 2  | l, 0 ) = H_{\mu^*_\phi, F}( \frac 1 2  | l, 1 ) = 0$. To sum, there exists $\bar p_{\mu^*_\phi, F} \in (0,1)$ such that for every $p < \bar p_{\mu^*_\phi, F}$, \eqref{eq:noninflumu} holds at prior state belief $\mu^*_\phi$. Because \eqref{eq:noninflumu} is strict and both sides of \eqref{eq:noninflumu} are continuous in $\mu$, the claim follows.
\end{proof}

Finally, because the left side of \eqref{eq:noninflumu} is continuous in $(\mu,p)$ and weakly continuous in $F$, because the right side of \eqref{eq:noninflumu} is continuous in $\mu$, and because \eqref{eq:noninflumu} is strict, part 2 follows.

\end{enumerate}

\subsection{Proof of \cref{thm:complementarity}}

The following lemma is essential.


\begin{lem*}\label{lem:primitive}
Fix $\mu \in [\frac 1 2, 1)$ and $F = (F^h, F^l)$ satisfying \cref{asp:lr2}. There exists $\gd_{\mu, F^l} > 0$ such that if $\max_{s \in S} \diff( F(\cdot|h,s), F^*(\cdot|s) ) < \gd_{\mu, F^l}$, then $\gb^\dagger_1(\mu, F) < \gb^\dagger_0(\mu, F)$ and so $F \in \mathcal F$.
\end{lem*}

\begin{proof}[Proof of \cref{lem:primitive}]
Fix $\mu \in [\frac 1 2, 1)$ and $F = (F^h, F^l)$ satisfying \cref{asp:lr2}. Note that as $F^h$ converges weakly to $F^*$, $\gb^\dagger_1(\mu, F^h, F^l)$ tends to $\ubar \gb^l(\mu, F^l)$ and $\gb^\dagger_0(\mu, F^h, F^l)$ tends to $\bar \gb^l(\mu, F^l)$. Because $\bar \gb^l(\mu, F^l) > \ubar \gb^l(\mu, F^l)$, the lemma follows. 
\end{proof}

Let $\Gamma(\xi)$ denote the game given information structure $\xi$. Fix $p \in (0,1)$. Let $(\mu^n)_{n=0}^\infty$ be a monotone sequence converging to 1, with $\mu^n \in [\frac 1 2, 1)$ for each $n$. Let $(F^k)_{k=0}^\infty$ be a sequence of experiment profiles along which the low type's experiment is constant and the high type's experiment converges weakly to $F^*$, and $F^k$ satisfies \cref{asp:lr2} for each $k$. In the following, for each $n$ and $k$, I write $\Gamma(\mu^n, p, F^k)$ as $\Gamma^{n,k}$. In $\Gamma^{n,k}$, I write the state belief distribution $H_{\mu^n, F^k}$ as $H^{n,k}$ and, for each state $s$, I write the crossing belief $\gb^\dagger_s(\mu^n, F^k)$ as $\gb^{\dagger,n,k}_s$. Finally, because $F^k(\cdot|l,\cdot)$ and so $H^{n,k}(\cdot|l,\cdot)$ is constant in $k$, I write $H^{n,k}(\cdot|l,\cdot)$ as $H^n(\cdot|l,\cdot)$ and the bounds $\ubar \gb^l$ and $\bar \gb^l$, defined in the main text, as $\ubar \gb^{n,l}$ and $\bar \gb^{n,l}$. I also write the bound $\bar \gb^h$ as $\bar \gb^{n, k, h}$, and write $\gb_{\mu^n, p, F^k}$ solving \eqref{eq:fixedpointbeta} as $\gb^{n,k}$. Note that by \cref{lem:primitive}, for each $n$, there exists $\ubar \gk^n>0$ such that if $k \ge \ubar \gk^n$, then $\gb^{\dagger,n,k}_1 < \gb^{\dagger,n,k}_0$ so that $F^k \in \mathcal F$ and so by \cref{prop:existence}, $\gb^{n,k}$ solving \eqref{eq:fixedpointbeta} is well-defined.

\begin{lem*}\label{lem:limithightypevaluable}
Let $(\ve^n)_{n=0}^\infty$ be a monotone sequence converging to zero, where $\ve^n \in (0,1-\mu^n)$ for each $n$. For each $n$, there exists $\gk^n > \ubar \gk^n$ such that if $k \ge \gk^n$, then given $\gb^{n,k}$ characterized according to \eqref{eq:fixedpointbeta} in $\Gamma^{n,k}$,
\begin{align*}
\bE^{\gb^{n,k}}[u(m,s)|h] > \mu^n + \ve^n.
\end{align*}
\end{lem*}

\begin{proof}[Proof of \cref{lem:limithightypevaluable}]
Fix $n$ and $\ve^n \in (0, 1-\mu^n)$. Because $ 0 < \ubar \gb^{n,l} < \bar \gb^{n,l} < 1$ and so
\begin{align*}
\lim_{k \ra \infty} H^{n,k}(\ubar \gb^{n,l}|h,0) &=1,\\
\lim_{k \ra \infty} H^{n,k}( \bar \gb^{n,l}|h,1) &=0,
\end{align*}
there exists $\gk^n>\ubar \gk^n$ such that for each $k \ge \gk^n$,
\begin{align}\label{eq:H0bound}
H^{n,k}(\ubar \gb^{n,l}|h,0) &> \mu^n + \ve^n,\\
\text{ and } \qquad  H^{n,k}( \bar \gb^{n,l}|h,1) &< 1 - \mu^n - \ve^n. \label{eq:H1bound}
\end{align}
Consequently, for each $k \ge \gk^n$, given $\gb^{n,k}$ characterized according to \cref{prop:existence},
\begin{align*}
\bE^{\gb^{n,k}}[u(m,s)|h] 
&=
(1-\mu^n) H^{n,k}(\gb^{n,k}|h,0)
+
\mu^n 
( 1 - H^{n,k}(\gb^{n,k}|h,1) ) \\
&> 
(1-\mu^n) H^{n,k}(\ubar \gb^{n,l}|h,0)
+
\mu^n 
( 1 - H^{n,k}(\bar \gb^{n,l} |h,1) )\\
&>
(1-\mu^n) (\mu^n + \ve^n) + \mu^n (\mu^n + \ve^n) \\
&= \mu^n  +  \ve^n,
\end{align*}
as desired, where the first inequality uses $\gb^{n,k} \in (\gb^{\dagger,n,k}_1, \gb^{\dagger,n,k}_0) \subsetneq (\ubar \gb^{n,l}, \bar \gb^{n,l})$ by \cref{lem:signals} and \cref{prop:existence}, and the second inequality uses \eqref{eq:H0bound} and \eqref{eq:H1bound}.
\end{proof}


\begin{lem*}\label{lem:mu2m2}
Let $(\ve^n)_{n=0}^\infty$ be a monotone sequence converging to zero, where $\ve^n \in (0,1-\mu^n)$ for each $n$. There exists $\bar n > 0$ such that for every $n \ge \bar n$, there exists $\tilde \gk^n> \ubar \gk^n$ such that for every $k \ge \tilde \gk^n$,
\begin{align}\label{eq:supineq}
H^n(\gb^{n,k}|l,1)  < \frac{\ve^n}{\mu^n} \frac{p}{1-p}.
\end{align}
\end{lem*}

\begin{proof}[Proof of \cref{lem:mu2m2}]
In \eqref{eq:fixedpointbeta}, by replacing $H^{n,k}(\gb|l,1)$ with  $\frac{\ve^n}{\mu^n} \frac{p}{1-p}$, the left side becomes
\begin{align*}
\textnormal{LS}^{n,k} := 
\begin{multlined}[t][12cm]
\gb^{n,k}  \frac{ p H^{n,k}(\gb^{n,k}|h,1) 
}
{
p H^{n,k}(\gb^{n,k}|h,1)
+
p \frac{\ve^n}{\mu^n}
} 
+ (1-\gb^{n,k}) \frac{p  H^{n,k}(\gb^{n,k}|h,0) 
}
{
p 
H^{n,k}(\gb^{n,k}|h,0)
+
(1-p) 
H^n(\gb^{n,k}|l,0) 
},
\end{multlined}
\end{align*}
and the right side becomes
\begin{align*}
\textnormal{RS}^{n,k} := 
\begin{multlined}[t][12.5cm]
\gb^{n,k} \frac{ p (1 - H^{n,k}(\gb^{n,k}|h,1))
}
{
p (1-H^{n,k}(\gb^{n,k}|h,1))
+
(1-p) \!\left( 1- \frac{\ve^n}{\mu^n} \frac{p}{1-p} \right)\!
}  
\\
+ (1-\gb^{n,k}) 
\frac{p (1- H^{n,k}(\gb^{n,k}|h,0) )
}
{
p 
(1 - H^{n,k}(\gb^{n,k}|h,0))
+
(1-p) 
(1 - H^n(\gb^{n,k}|l,0) )
}.
\end{multlined}  
\end{align*}
Because \eqref{eq:fixedpointbeta} holds in any informative equilibrium $\gb$, and because the left side of \eqref{eq:fixedpointbeta} is strictly decreasing in $H_{\mu,F}(\gb|l,1)$ and the right side is strictly increasing in $H_{\mu,F}(\gb|l,1)$, to prove that \eqref{eq:supineq} holds for sufficiently large $n$ and sufficiently large $k$ relative to $n$, it suffices to show $\lim_{n \ra \infty} \lim_{k \ra \infty} \textnormal{LS}^{n,k} < \lim_{n \ra \infty} \lim_{k \ra \infty} \textnormal{RS}^{n,k}$. Because
$$1 \ge \lim_{n \ra \infty} \lim_{k \ra \infty} \gb^{n,k} \ge \lim_{n \ra \infty} \lim_{k \ra \infty} \ubar \gb^{n,l} = \lim_{n \ra \infty}  \ubar \gb^{n,l} = 1,$$
it  holds that $\lim_{n \ra \infty} \lim_{k \ra \infty} \gb^{n,k} = 1$. Moreover, $\lim_{k \ra \infty} H^{n,k}(\gb^{n,k}|h,1) =0$ because
\begin{align*}
0 \le \lim_{k \ra \infty} H^{n,k}(\gb^{n,k}|h,1) \le \lim_{k \ra \infty} H^{n,k}(\bar \gb^{n,l}|h,1) = 0,
\end{align*}
where the latter inequality holds because $\gb^{n,k} < \gb^{\dagger,n,k}_0 < \bar \gb^{n,l}$ for each $k$ by \cref{lem:signals} and \cref{prop:existence}, and the equality follows because $\bar \gb^{n,l} < 1$. Consequently,
\begin{align*}
\lim_{n \ra \infty} \lim_{k \ra \infty} \textnormal{LS}^{n,k} = 0 < p = \lim_{n \ra \infty} \lim_{k \ra \infty} \textnormal{RS}^{n,k},
\end{align*}
as was to be shown. 
\end{proof}

For each $n$ and $k \ge \ubar \gk_n$, given $\gb^{n,k}$ characterized by \eqref{eq:fixedpointbeta} in $\Gamma^{n,k}$, write the receiver's matching payoff in informative equilibrium $(\gb^{n,k}, \tau)$ conditional on type $t$ as
\begin{align}\label{eq:mainEu}
\begin{multlined}[t][14cm]
U^{n,k}(\gb^{n,k}|t)
:= \mu^n [
1 - H^{n,k}(\gb^{n,k}|t,1) 
]
+
(1-\mu^n) H^{n,k}(\gb^{n,k}|t,0).
\end{multlined}
\end{align}
By \cref{lem:limithightypevaluable} and \cref{lem:mu2m2}, for each $n \ge \bar n$, there is $\bar k^n \ge \max[\gk^n, \tilde \gk^n]$, where $\gk^n \ge \ubar \gk^n$ and $\tilde \gk^n \ge \ubar \gk^n$ are identified above, such that for every $k \ge \bar k^n$,
\begin{align}\label{eq:Unkhbd}
U^{n,k}(\gb^{n,k}|h)  > \mu^n + \ve^n,
\end{align}
and
\begin{align}\nonumber
U^{n,k}(\gb^{n,k}|l) &= 
\begin{multlined}[t][12cm]
\mu^n
[
1 - H^{n,k}(\gb^{n,k}|l,1) 
]
+
(1-\mu^n)
[
H^{n,k}(\gb^{n,k}|l,0) 
]
\end{multlined}\\ \nonumber
&\ge \mu^n
[
1 - H^{n,k}(\gb^{n,k}|l,1) 
] \\
&> \mu^n - \ve^n \frac{p}{1-p}.
\label{eq:Unklbd}
\end{align}
where the last line uses \eqref{eq:supineq}. Thus, \eqref{eq:advicequality} holds, namely $p U^{n,k}(\gb^{n,k}|h) + (1-p)U^{n,k}(\gb^{n,k}|l) > \mu^n$, because
\begin{align*}
&~p U^{n,k}(\gb^{n,k}|h) + (1-p)U^{n,k}(\gb^{n,k}|l) - \mu^n \\
=&~ 
p (
U^{n,k}(\gb^{n,k}|h) - \mu^n
)
+
(1-p)
(
U^{n,k}(\gb^{n,k}|l) - \mu^n
)
\\
>&~ p \ve^n - p \ve^n = 0,
\end{align*}
as desired, where the last line uses \eqref{eq:Unkhbd} and \eqref{eq:Unklbd}.  

\subsection{Proof of \cref{prop:highnoninflu}}

Fix $(p, F) \in (0,1) \times \mathcal F$. At $\mu = 1/(1+\ubar \ell^h)$, the high type's state belief is at least $\frac 1 2$ irrespective of her signal. Because $\ubar \ell^l > \ubar \ell^h$ by \cref{asp:lr2}, the low type's state belief is also at least $\frac 1 2$ irrespective of her signal. Therefore, if each type truthfully reports the state she deems likely, the receiver's matching payoff is
\begin{align*}
\bE^p \!\left[ \frac{1}{1+\ubar \ell^h} \!\left( 1- H_{1/(1+\ubar \ell^h),F} \!\left( \frac 1 2 \middle| t, 1 \right)\! \right)\!  + \frac{\ubar \ell^h}{1+\ubar \ell^h}  H_{1/(1+\ubar \ell^h),F} \!\left( \frac 1 2 \middle| t, 0 \right)\! \right] = \frac{1}{1+\ubar \ell^h}.
\end{align*}
Therefore, if $\gb_{1/(1+\ubar \ell^h),p,F} > \frac 1 2$, then the receiver's matching payoff is
\begin{align*}
&~ \bE^p \!\left[ \frac{1}{1+\ubar \ell^h} \!\left( 1- H_{1/(1+\ubar \ell^h),F} \!\left( \gb_{1/(1+\ubar \ell^h),p,F} \middle| t, 1 \right)\! \right)\!  + \frac{\ubar \ell^h}{1+\ubar \ell^h}  H_{1/(1+\ubar \ell^h),F} \!\left( \gb_{1/(1+\ubar \ell^h),p,F} \middle| t, 0 \right)\! \right] \\
<&~ \bE^p \!\left[ \frac{1}{1+\ubar \ell^h} \!\left( 1- H_{1/(1+\ubar \ell^h),F} \!\left( \frac 1 2 \middle| t, 1 \right)\! \right)\!  + \frac{\ubar \ell^h}{1+\ubar \ell^h}  H_{1/(1+\ubar \ell^h),F} \!\left( \frac 1 2 \middle| t, 0 \right)\! \right] \\
=&~ \frac{1}{1+\ubar \ell^h},
\end{align*}
where the inequality follows from \cref{lem:maxat12}. By continuity, for any $\mu$ in the neighborhood of $1/(1+\ubar \ell^h)$, if $(\mu, p, F) \in \Xi^*$ so that $\mu \in  (\mu^I_{p,F}, 1/(1+\ubar \ell^h))$ and therefore $\gb_{\mu,p,F} > \frac 1 2$, the receiver's matching payoff is strictly lower than his reservation payoff:
\begin{align*}
\bE^p [ \mu ( 1- H_{\mu,F} ( \gb_{\mu,p,F} | t, 1 ) )  + (1-\mu)  H_{\mu,F} ( \gb_{\mu,p,F} | t, 0 ) ] < \mu.
\end{align*}
Consequently, the proposition follows.



\subsection{Proof of \cref{prop:substitute}}

By \cref{prop:existence}, there exists a unique $\gb_{\mu,p,F} \in (\gb^\dagger_1(\mu, F), \gb^\dagger_0(\mu, F))$ satisfying \eqref{eq:fixedpointbeta} for each $\mu \in [\frac 1 2, 1)$. It remains to show that given such $\gb_{\mu, p,F}$, \eqref{eq:advicequality} holds if $\mu$ is sufficiently close to $\frac 1 2$. Consider first $\mu = \frac 1 2$. For each type $t$, consider the inequality
\begin{align*}
\frac 1 2 H_{\mu, F}(\gb_{\frac 1 2, p,F}|t,0) + \frac 1 2 
(1-H_{\mu, F}(\gb_{\frac 1 2, p,F}|t,1))  > \frac 1 2.
\end{align*}
This inequality holds because it can be written as $H_{\mu, F}(\gb_{\frac 1 2, p,F}|t,0) > H_{\mu, F}(\gb_{\frac 1 2, p,F}|t,1)$, which holds by \cref{lem:mlrpimplication}. Therefore, \eqref{eq:advicequality} holds strictly at $\mu=\frac 1 2$. By continuity of the left side of  \eqref{eq:advicequality} in $\mu$, the proposition follows.

\subsection{Proof of \cref{prop:compare}}

Call the pair $\pi := (p, F)$ a signal structure, and let $\Pi := (0,1) \times \mathcal F$ denote the set of all signal structures where the experiment profile belongs to $\mathcal F$.

Part 2 of \cref{prop:noninflusource} implies that there is a set of signal structures $\tilde \Pi^*$, with nonempty interior, in $\Pi$ such that for each $\pi \in \tilde \Pi^*$, there is a prior state belief $\mu \in [\frac 1 2, 1)$ such that the resulting $\gb_{\mu,\pi}$-cutoff strategy profile satisfies $\bE^{\gb_{\mu,\pi}}[u(m,s)] < \mu$, and the informative equilibrium is noninfluential. Fixing any such $\pi \in \tilde \Pi^*$, \cref{prop:substitute} ensures that there exists $\ubar \mu_{\pi} > \frac 1 2$ such that for each $\mu \in [\frac 1 2, \ubar \mu_{\pi})$, the informative equilibrium is influential, with $\bE^{\gb_{\mu,\pi}}[u(m,s)] > \mu$. Consequently, fixing $\pi \in \tilde \Pi^*$, by continuity of $\bE^{\gb_{\mu,\pi}}[u(m,s)]$ in $\mu$, there exist $\mu'_\pi, \mu''_\pi \in [\frac 1 2, 1)$ satisfying $\mu''_\pi < \mu'_\pi$ such that:
\begin{enumerate}\itemsep0em
\item $\bE^{\gb_{\mu''_\pi,\pi}}[u(m,s)] > \mu''_\pi$;
\item $\bE^{\gb_{\mu'_\pi,\pi}}[u(m,s)] < \mu'_\pi$;
\item there is a unique $\tilde \mu_\pi \in (\mu''_\pi, \mu'_\pi)$ such that $\bE^{\gb_{\tilde \mu_\pi,\pi}}[u(m,s)] = \tilde \mu_\pi$;
\item $\mu''_\pi$ is sufficiently close to $\tilde \mu_\pi$ such that $\bE^{\gb_{\mu''_\pi,\pi}}[u(m,s)]$ is sufficiently close to $\tilde \mu_\pi$ and in turn $\bE^{\gb_{\mu''_\pi,\pi}}[u(m,s)] < \mu'_\pi$.
\end{enumerate}
Finally, \cref{thm:complementarity} implies that there is an open subset $\Pi^*$ of $\tilde \Pi^*$ such that for each $\pi \in \Pi^*$, there exists $\mu'''_\pi > \mu'_\pi$ such that $\bE^{\gb_{\mu'''_\pi,\pi}}[u(m,s)] > \mu'''_\pi$ and the informative equilibrium is influential. Consequently, for each $\pi \in \Pi^*$, the informative equilibrium is influential when the prior state belief is equal to either $\mu''_\pi$ or $\mu'''_\pi$ and the informative equilibrium is noninfluential at prior state belief $\mu'_\pi$. Finally, because $U_\pi(\mu)=\max(\bE^{\gb_{\mu,\pi}}[u(m,s)], \mu)$, it follows that
\begin{align*}
U_\pi(\mu''_\pi) = \bE^{\gb_{\mu''_\pi,\pi}}[u(m,s)]  < \mu'_\pi = U_\pi(\mu'_\pi) < \mu'''_\pi < \bE^{\gb_{\mu'''_\pi,\pi}}[u(m,s)] = U_\pi(\mu'''_\pi),
\end{align*}
as desired.

\subsection{Proof of \cref{thm:highermuhigherpayoff}}

Fix any sequence of experiment profiles $(F^k)_{k=0}^\infty$ as stated. To ease the notation, because $F^k(\cdot|l,\cdot)$ and thus $f^k(\cdot|l,\cdot)$ are independent of $k$, I write them as $F(\cdot|l,\cdot)$ and $f(\cdot|l,\cdot)$ when it is convenient. I also write $\gb_{\mu, p, F^k}$ as $\gb^k_\mu$. The arguments in the proofs of \cref{prop:existence} extend to the limit as $k \ra \infty$, so that $\gb^\infty_\mu := \lim_{k \ra \infty} \gb^k_\mu$ exists. By the Portmanteau theorem (\egg \citealp[Theorem 2.1]{billingsley2013convergence}), weak convergence of the high type's experiment to $F^*$ is equivalent to the high type's signal densities satisfying, as $k \to \infty$, $f^k(\cdot|h,0) \to 0 \text{ in } L^1_{\textnormal{loc}}((0, \infty])$ and $f^k(\cdot|h,1) \to 0 \text{ in } L^1_{\textnormal{loc}}([0, \infty))$, because $F^*(\cdot|s)$ is a singular distribution for each state $s$.

For each $k$, the receiver's informative-equilibrium matching payoff is
\begin{align}\nonumber
\bE^{\gb^k_\mu}[u(m,s)] =&~ \bE^p[\mu (1-H_{\mu, F^k}(\gb^k_\mu|t,1)) + (1-\mu) H_{\mu, F^k}(\gb^k_\mu|t,0) )] \\
=&~ \bE^p[ \mu (1-F(\ell^k_\mu|t,1)) + (1-\mu) F(\ell^k_\mu|t,0) )], \label{eq:matchingpayoffbeta0}
\end{align}
where
\begin{align}\label{eq:ellkmu}
\ell^k_\mu := \frac{1-\mu}{\mu} \frac{\gb^k_\mu}{1-\gb^k_\mu}.
\end{align}
Write $\ell^\infty_\mu := \lim_{k \ra \infty} \ell^k_\mu$. For each $k$, the derivative of \eqref{eq:matchingpayoffbeta0} with respect to $\mu$ is
\begin{align}\label{eq:derivativemu}
\begin{multlined}[12cm]
p \!\left( 1 - F^k(\ell^k_\mu|h,0) - F^k(\ell_\mu|h,1) + \frac{\diff \ell^k_\mu}{\diff \mu}  ( (1-\mu)f^k(\ell^k_\mu|h,0)  -  \mu f^k(\ell^k_\mu|h,1) ) \right) \\
+ 
(1-p) \!\left( 1 - F(\ell^k_\mu|l,0) - F(\ell^k_\mu|l,1) + \frac{\diff \ell^k_\mu}{\diff \mu}  ( (1-\mu)f(\ell^k_\mu|l,0)  -  \mu f(\ell^k_\mu|l,1) ) \right).
\end{multlined}
\end{align}
For each $k$ and $\mu$, 
$\ell^k_\mu \in (\ubar \ell^l, \bar \ell^l)$ because $\gb^k_\mu \in (\ubar \gb^l, \bar \gb^l)$. Because $(\ubar \gb^l, \bar \gb^l) \subsetneq (\ubar \gb^h, \bar \gb^h)$ by \cref{asp:lr2}(a), $\ell^\infty_\mu \in (\ubar \ell^h, \bar \ell^h)$. Because $f^k(\cdot|h,0) \ra 0$ in $L^1_{\textnormal{loc}}((0, \infty])$ and $f^k(\cdot|h,1) \ra 0$ in $L^1_{\textnormal{loc}}([0, \infty))$, $F^k$ converges weakly to $F^*$. Therefore, as $k \ra \infty$,
\begin{align}\label{eq:lim1}
1 - F^k(\ell^k_\mu|h,0) - F^k(\ell^k_\mu|h,1) &\ra 0,\\ \label{eq:lim2}
f^k(\ell^k_\mu|h,0) &\ra 0,\\ \label{eq:lim3}
f^k(\ell^k_\mu|h,1) &\ra 0.
\end{align}
By \eqref{eq:ellkmu}, the equilibrium condition \eqref{eq:fixedpointbetanew} can be written as
\begin{align}\label{eq:ellstarappnew}
\begin{multlined}[12cm]
\frac{\ell^k_\mu \mu}{\ell \mu + 1 - \mu} \bigg[
\frac{p (1-F^k(\ell^k_\mu|h,1))}{
\bE^p[1-F^k(\ell^k_\mu|t,1)]} 
-
\frac{p F^k(\ell^k_\mu|h,1)}{\bE^p[ F^k(\ell^k_\mu|t,1)]}
\bigg] 
\\[0.25em]
= \frac{1-\mu}{\ell \mu + 1 - \mu}  \bigg[
\frac{p F^k( \ell^k_\mu |h,0) }{\bE^p[F^k( \ell^k_\mu |t,0)]
}
-
\frac{p (1-F^k(\ell^k_\mu|h,0))}{\bE^p[1-F^k(\ell^k_\mu|t,0)]} 
\bigg].
\end{multlined}
\end{align}
Following the same arguments in the proof of \cref{prop:existence}, the left side of \eqref{eq:ellstarappnew} is strictly increasing in $\ell^k_\mu$ and, holding $\ell^k_\mu$ fixed, is also strictly increasing in $\mu$ , whereas the right side of \eqref{eq:ellstarappnew} is strictly decreasing in $\ell^k_\mu$ and, holding $\ell^k_\mu$ fixed, is also strictly decreasing in $\mu$. Consequently, $\ell^k_\mu$ solving \eqref{eq:ellstarappnew} satisfies
\begin{align*}
\frac{\diff \ell^k_\mu}{\diff \mu} < 0.
\end{align*}
I prove the two parts of the theorem in order.
\begin{enumerate}
\item Fix $\mu \in [\frac 1 2, \mu^I_p)$. The arguments in \cref{prop:cutoff} extend to the limit as $k \ra \infty$, giving $\gb^\infty_\mu < \frac 1 2$, or equivalently, 
\begin{align}\label{eq:fsign0}
(1-\mu)f(\ell^\infty_\mu|l,0) - \mu f(\ell^\infty_\mu|l,1) > 0.
\end{align}
It also holds that 
\begin{align}\label{eq:Fsign0}
1-F( \ell^\infty_\mu |l,1) < F( \ell^\infty_\mu |l,0).   
\end{align}
To see this, if towards a contradiction $1-F( \ell^\infty_\mu |l,1) \ge F( \ell^\infty_\mu |l,0)$, then
\begin{align*}
\frac{\gb^\infty_\mu}{p + (1-p) (1 - F( \ell^\infty_\mu |l,1)) } &<
\frac{\frac 1 2}{p + (1-p) (1 - F( \ell^\infty_\mu |l,1)) } \\
&\le \frac{\frac 1 2}{p + (1-p) F( \ell^\infty_\mu |l,0) } \\
&< \frac{1-\gb^\infty_\mu}{p + (1-p) F(\ell^\infty_\mu |l,0) },
\end{align*}
but by \eqref{eq:lim1}---\eqref{eq:lim3}, the equilibrium condition \eqref{eq:fixedpointbeta} in the limit as $k \ra \infty$ simplifies to
\begin{align}\label{eq:limeqm}
\frac{\gb^\infty_\mu}{p + (1-p) (1 - F( \ell^\infty_\mu |l,1)) } = \frac{1-\gb^\infty_\mu}{p + (1-p) F( \ell^\infty_\mu |l,0) },
\end{align}
a contradiction. By \eqref{eq:lim1}---\eqref{eq:lim3}, as $k \ra \infty$, the derivative \eqref{eq:derivativemu} tends to
\begin{align*}
\begin{multlined}[t][12cm]
(1-p) \bigg( \underbrace{ \vphantom{\frac{\diff \ell^\infty_\mu}{\diff \mu}} 1 - F(\ell^\infty_\mu|l,0) - F(\ell^\infty_\mu|l,1)}_{\textnormal{<0, by \eqref{eq:Fsign0}}} 
+ \underbrace{ \vphantom{\frac{\diff \ell^\infty_\mu}{\diff \mu}} \!\left( \lim_{k \ra \infty}\frac{\diff \ell^k_\mu}{\diff \mu} \right)\!}_{<0} \underbrace{\vphantom{\frac{\diff \ell^\infty_\mu}{\diff \mu}} ( (1-\mu)f(\ell^\infty_\mu|l,0)  -  \mu f(\ell^\infty_\mu|l,1) )}_{\textnormal{>0, by \eqref{eq:fsign0}}} \bigg) < 0.
\end{multlined}
\end{align*}
Consequently, for sufficiently large $k$, a marginal increase in $\mu$ strictly decreases the receiver's informative-equilibrium matching payoff. Moreover, by part 1 of \cref{prop:noninflusource}, because $\mu < \mu^I_{p,F^k}$ for sufficiently large $k$, $\gb^k_\mu < \frac 1 2$ and so the receiver's informative-equilibrium payoff is equal to his informative-equilibrium matching payoff. Therefore a marginal increase in $\mu$ also strictly decreases the receiver's informative-equilibrium payoff.
\item Fix $\mu \in (\mu^I_p, 1)$. The arguments in \cref{prop:cutoff} extend to the limit as $k \ra \infty$, giving $\gb^\infty_\mu > \frac 1 2$, or equivalently,
\begin{align}\label{eq:fsign}
(1-\mu)f(\ell^\infty_\mu|l,0) - \mu f(\ell^\infty_\mu|l,1) < 0.
\end{align}
It also holds that 
\begin{align}\label{eq:Fsign}
1-F( \ell^\infty_\mu |l,1) > F( \ell^\infty_\mu |l,0).   
\end{align}
To see this, if towards a contradiction $1-F( \ell^\infty_\mu |l,1) \le F( \ell^\infty_\mu |l,0)$, then
\begin{align*}
\frac{\gb^\infty_\mu}{p + (1-p) (1 - F( \ell^\infty_\mu |l,1)) } &>
\frac{\frac 1 2}{p + (1-p) (1 - F( \ell^\infty_\mu |l,1)) } \\
&\ge \frac{\frac 1 2}{p + (1-p) F( \ell^\infty_\mu |l,0) } \\
&> \frac{1-\gb^\infty_\mu}{p + (1-p) F(\ell^\infty_\mu |l,0) },
\end{align*}
but by \eqref{eq:lim1}---\eqref{eq:lim3}, the equilibrium condition \eqref{eq:fixedpointbeta} in the limit as $k \ra \infty$ simplifies to \eqref{eq:limeqm}, a contradiction. By \eqref{eq:lim1}---\eqref{eq:lim3}, as $k \ra \infty$, the derivative \eqref{eq:derivativemu} tends to
\begin{align*}
\begin{multlined}[t][12cm]
(1-p) \bigg( \underbrace{ \vphantom{\frac{\diff \ell^\infty_\mu}{\diff \mu}} 1 - F(\ell^\infty_\mu|l,0) - F(\ell^\infty_\mu|l,1)}_{\textnormal{>0, by \eqref{eq:Fsign}}} 
+ \underbrace{ \vphantom{\frac{\diff \ell^\infty_\mu}{\diff \mu}} \!\left( \lim_{k \ra \infty}\frac{\diff \ell^k_\mu}{\diff \mu} \right)\!}_{<0} \underbrace{\vphantom{\frac{\diff \ell^\infty_\mu}{\diff \mu}} ( (1-\mu)f(\ell^\infty_\mu|l,0)  -  \mu f(\ell^\infty_\mu|l,1) )}_{\textnormal{<0, by \eqref{eq:fsign}}} \bigg) > 0.
\end{multlined}
\end{align*}
Consequently, for sufficiently large $k$, a marginal increase in $\mu$ strictly decreases the receiver's informative-equilibrium matching payoff. Moreover, this also strictly increases the receiver's informative-equilibrium payoff, because this latter payoff is a maximum between the matching payoff and the receiver's reservation $\mu$.
\end{enumerate}

\subsection{Proof of \cref{prop:alloeffrep}}

Fix any sequence of experiment profiles $(F^k)_{k=0}^\infty$ as stated. By the Portmanteau theorem (\egg \citealp[Theorem 2.1]{billingsley2013convergence}), weak convergence of the high type's experiment to $F^*$ is equivalent to the high type's signal densities satisfying, as $k \to \infty$, $f^k(\cdot|h,0) \to 0 \text{ in } L^1_{\textnormal{loc}}((0, \infty])$ and $f^k(\cdot|h,1) \to 0 \text{ in } L^1_{\textnormal{loc}}([0, \infty))$, because $F^*(\cdot|s)$ is a singular distribution for each state $s$. Let $\mu^I_p := \lim_{k\to\infty}\mu^I_{p,F^k}$, where $\mu^I_{p,F^k}$ is identified in \cref{prop:cutoff}. For each $k$, let $\ubar \ell^{k,h}$ denote the lower bound of the support of the high type's signal. I prove the two parts of the proposition in order.
\begin{enumerate}
\item For each $k$, $I_{p,F^k} \subsetneq [\frac 1 2, 1/(1+\ubar \ell^{k,h}))$ by \cref{prop:highnoninflu}. Part 1 then follows by showing that for sufficiently large $k$, the receiver's informative-equilibrium matching payoff is strictly increasing in $p$. Write $\gb_{\mu, p, F^k}$ simply as $\gb^k_p$, and write
\begin{align}\label{eq:ellkp}
\ell^k_p := \frac{1-\mu}{\mu} \frac{\gb^k_p}{1-\gb^k_p}.
\end{align}
The receiver's informative-equilibrium matching payoff can be written as
\begin{align}
\bE^{\gb^k_p}[u(m,s)] = \bE^p[\mu (1-F(\ell^k_p|t,1)) + (1-\mu) F(\ell^k_p|t,0)]. \label{eq:matchingpayoffbetap}
\end{align}
The derivative of \eqref{eq:matchingpayoffbetap} with respect to $p$ is
\begin{align}\label{eq:derivativep}
\begin{multlined}[13cm]
\overbrace{\left[\mu (1-F^k(\ell^k_p|h,1)) + (1-\mu) F^k(\ell^k_p|h,0) \right]- \left[ \mu (1-F(\ell^k_p|l,1)) + (1-\mu) F(\ell^k_p|l,0) \right]}^{\textnormal{term A}} \\
+ \underbrace{\frac{\diff \ell^k_p}{\diff p} \bE^p\left[ (1-\mu) f^k(\ell^k_p|t,0) -\mu f^k(\ell^k_p|t,1) \right]}_{\textnormal{term B}}.
\end{multlined}
\end{align}
For each $k$, because $\gb^k_p \in (\gb^{\dagger,k}_1, \gb^{\dagger,k}_0)$, $1-H(\gb^k_p|h,1) > 1-H(\gb^k_p|l,1)$ and $H(\gb^k_p|h,0) > H(\gb^k_p|l,0)$ by \cref{lem:signals}. In the limit as $k \ra \infty$, term A tends to
\begin{align*}
1 - \left[ \mu (1-F(\ell^\infty_p|l,1)) + (1-\mu) F(\ell^\infty_p|l,0) \right] > 0.
\end{align*}
\cref{lem:dldp} turns to term B and completes the proof of part 1.
\begin{lem*}\label{lem:dldp}
It holds that
\begin{align*}
\lim_{k \ra \infty} \frac{\diff \ell^k_p}{\diff p} \bE^p\!\left[ (1-\mu) f(\ell^k_p|t,0) -\mu f(\ell^k_p|t,1) \right] ~\begin{cases}
~> 0, \quad &\text{ if } 1 - F(\ell^\infty_p|l,1) \neq F(\ell^\infty_p|l,0),\\
~= 0, \quad &\text{ if } 1 - F(\ell^\infty_p|l,1) = F(\ell^\infty_p|l,0).\\
\end{cases}
\end{align*}
\end{lem*}

\begin{proof}[Proof of \cref{lem:dldp}]
Rewrite the equilibrium condition \eqref{eq:fixedpointbeta} as
\begin{align}
\gb^k_p A^k_1(\gb^k_p,p) - (1-\gb^k_p)A^k_0(\gb^k_p,p) = 0.
\label{eq:Psi}
\end{align}
where
\begin{align*}
A^k_1(\gb,p) &:=
\frac{1-F^k(\gb|h,1)}{p(1-F^k(\gb|h,1))+(1-p)(1-F^k(\gb|l,1))}
-
\frac{F^k(\gb|h,1)}{p F^k(\gb|h,1)+(1-p) F^k(\gb|l,1)}, \\
A_0(\gb,p) &:=
\frac{F^k(\gb|h,0)}{p F^k(\gb|h,0) +(1-p) F^k(\gb|l,0)}
-
\frac{1-F^k(\gb|h,0)}{p(1-F^k(\gb|h,0))+(1-p)(1-F^k(\gb|l,0))}.
\end{align*}
For each state $s$, write the crossing belief $\gb^\dagger_s(\mu, F^k)$ as $\gb^{\dagger,k}_s$. Because $\gb^k_p \in (\gb^{\dagger,k}_1, \gb^{\dagger,k}_0)$, $F^k(\gb^k_p|h,1) < F^k(\gb^k_p|l,1)$ by \cref{lem:signals} and so $A^k_1(\gb^k_p,p) > 0$. As shown in the proof of \cref{prop:existence}, the left side of \eqref{eq:Psi} is strictly increasing in $\gb$. Therefore, by the implicit function theorem, the sign of $\diff \gb^k_p/\diff p$ is the negative of the sign of the derivative of the left side of \eqref{eq:Psi} with respect to $p$, holding $\gb^k_p$ fixed. This derivative is
\begin{align*}
\gb^k_p \partial A^k_1(\gb^k_p,p)-(1-\gb^k_p) \partial_p A^k_0(\gb^k_p,p), 
\end{align*}
where 
\[
\partial_p A^k_1
=
\begin{multlined}[t][13cm]
A^k_1(\gb^k_p,p) \bigg(
-\frac{1}{1-p}
+\frac{F^k(\ell^k_p|l,1)-F^k(\ell^k_p|h,1)}{p F^k(\ell^k_p|h,1) +(1-p) F^k(\ell^k_p|l,1)}
\\ -\frac{F^k(\ell^k_p|l,1) - F^k(\ell^k_p|h,1)}{p(1-F^k(\ell^k_p|h,1))+(1-p)(1-F^k(\ell^k_p|l,1))}
\bigg),
\end{multlined}
\]
\[
\partial_p A^k_0
=
\begin{multlined}[t][13cm]
A^k_0(\gb^k_p,p) \bigg(
-\frac{1}{1-p}
+\frac{F^k(\ell^k_p|h,0) - F^k(\ell^k_p|l,0)}{p(1-F^k(\ell^k_p|h,0))+(1-p)(1-F^k(\ell^k_p|l,0))}
\\ -\frac{F^k(\ell^k_p|h,0)-F^k(\ell^k_p|l,0)}{p F^k(\ell^k_p|h,0) + (1-p)F^k(\ell^k_p|l,0)}
\bigg).
\end{multlined}
\]
Using \eqref{eq:Psi}, it follows that
\begin{align}\nonumber
&~ \gb^k_p \partial A^k_1(\gb^k_p,p)-(1-\gb^k_p) \partial_p A^k_0(\gb^k_p,p)
\\
=&~
\gb^k_p A_1(\gb^k_p,p) \Gamma(\gb^k_p,p),
\label{eq:Psi_p}
\end{align}
where
\begin{align*}
&~ \Gamma^k(\gb^k_p, p) \\
:=&~ 
\begin{multlined}[t][13cm]
\left(
\frac{F^k(\ell^k_p|l,1) - F^k(\ell^k_p|h,1)}{p F^k(\ell^k_p|h,1)+(1-p)F^k(\ell^k_p|l,1)}
-\frac{F^k(\ell^k_p|l,1) - F^k(\ell^k_p|h,1)}{p (1-F^k(\ell^k_p|h,1))+(1-p)(1-F^k(\ell^k_p|l,1))}
\right)\\
-
\left(
\frac{ F^k(\ell^k_p|h,0) - F^k(\ell^k_p|l,0)}{p(1-F^k(\ell^k_p|h,0))+(1-p)(1-F^k(\ell^k_p|l,0))}
-\frac{ F^k(\ell^k_p|h,0) - F^k(\ell^k_p|l,0) }{p F^k(\ell^k_p|h,0)  +(1-p) F^k(\ell^k_p|l,0) }
\right).
\end{multlined}
\end{align*}
Since $\gb^k_p A^k_1(\gb^k_p,p)>0$, the sign of \eqref{eq:Psi_p} equals the sign of $\Gamma(\gb^k_p, p)$. By direct computation,
\begin{align*}
\lim_{k \ra \infty} \Gamma^k(\gb^k_p, p)
&=
\frac{1- F(\ell^\infty_p|l,0)}{p + (1-p)F(\ell^\infty_p|l,0)} -\frac{F(\ell^\infty_p|l,1)}{p + (1-p)(1-F(\ell^\infty_p|l,1))}  \\
&~\begin{cases}
~< 0, \quad &\text{ if } 1 - F(\ell^\infty_p|l,1) < F(\ell^\infty_p|l,0),\\
~= 0, \quad &\text{ if } 1 - F(\ell^\infty_p|l,1) = F(\ell^\infty_p|l,0),\\
~> 0, \quad &\text{ if } 1 - F(\ell^\infty_p|l,1) > F(\ell^\infty_p|l,0).
\end{cases}
\end{align*}
Because the sign of $\diff \gb^k_p/\diff p$ is the negative of that of $\Gamma(\gb^k_p, p)$, by \eqref{eq:ellkp},
\begin{align*}
\lim_{k \ra \infty} \frac{\diff \ell^k_p}{\diff p} = \lim_{k \ra \infty} \frac{\diff \gb^k_p}{\diff p} \underbrace{\frac{(1-\mu)(1-p)}{\mu (1-\gb^k_p)^2}}_{ > 0} ~\begin{cases}
~> 0, \quad &\text{ if } 1 - F(\ell^\infty_p|l,1) < F(\ell^\infty_p|l,0),\\
~= 0, \quad &\text{ if } 1 - F(\ell^\infty_p|l,1) = F(\ell^\infty_p|l,0),\\
~< 0, \quad &\text{ if } 1 - F(\ell^\infty_p|l,1) > F(\ell^\infty_p|l,0).
\end{cases}
\end{align*}
Next, following the same arguments as in the proof of \cref{thm:highermuhigherpayoff}, for each $\mu < \mu^I_p$, $\gb^\infty_p < \frac 1 2$, $(1-\mu)f(\ell^\infty_p|l,0) - \mu f(\ell^\infty_p|l,1) > 0$ and $1-F( \ell^\infty_p |l,1) < F( \ell^\infty_p |l,0)$. Consequently, 
\begin{align*}
&~ \lim_{k \ra \infty} \frac{\diff \ell^k_p}{\diff p} \bE^p\!\left[ (1-\mu) f^k(\ell^k_p|t,0) -\mu f^k(\ell^k_p|t,1) \right] \\
=&~ \bigg( \underbrace{\lim_{k \ra \infty} \frac{\diff \ell^k_p}{\diff p}}_{>0} \bigg) \underbrace{ \vphantom{\lim_{k \ra \infty} \frac{\diff \ell^k_p}{\diff p}} (1-p)( (1-\mu) f(\ell^\infty_p|l,0) -\mu f(\ell^\infty_p|l,1) )}_{ > 0} > 0.
\end{align*}
For $\mu = \mu^I_p$, $\gb^\infty_p = \frac 1 2$, $(1-\mu)f(\ell^\infty_p|l,0) - \mu f(\ell^\infty_p|l,1) = 0$ and $1-F( \ell^\infty_\mu |l,1) = F( \ell^\infty_\mu |l,0)$. Consequently, 
\begin{align*}
&~ \lim_{k \ra \infty} \frac{\diff \ell^k_p}{\diff p} \bE^p\!\left[ (1-\mu) f^k(\ell^k_p|t,0) -\mu f^k(\ell^k_p|t,1) \right] \\
=&~ \bigg( \underbrace{\lim_{k \ra \infty} \frac{\diff \ell^k_p}{\diff p}}_{ = 0} \bigg) \underbrace{ \vphantom{\lim_{k \ra \infty} \frac{\diff \ell^k_p}{\diff p}} (1-p)( (1-\mu) f(\ell^\infty_p|l,0) -\mu f(\ell^\infty_p|l,1) )}_{ = 0} = 0.
\end{align*}
Finally, for each $\mu > \mu^I_p$, $\gb^\infty_p > \frac 1 2$, $(1-\mu)f(\ell^\infty_p|l,0) - \mu f(\ell^\infty_p|l,1) < 0$ and $1-F( \ell^\infty_\mu |l,1) > F( \ell^\infty_\mu |l,0)$. Consequently, 
\begin{align*}
&~ \lim_{k \ra \infty} \frac{\diff \ell^k_p}{\diff p} \bE^p\!\left[ (1-\mu) f^k(\ell^k_p|t,0) -\mu f^k(\ell^k_p|t,1) \right] \\
=&~ \bigg( \underbrace{\lim_{k \ra \infty} \frac{\diff \ell^k_p}{\diff p}}_{<0} \bigg) \underbrace{ \vphantom{\lim_{k \ra \infty} \frac{\diff \ell^k_p}{\diff p}} (1-p)( (1-\mu) f(\ell^\infty_p|l,0) -\mu f(\ell^\infty_p|l,1) )}_{ < 0} > 0.
\end{align*}
This completes the proof.
\end{proof}
\item The receiver's informative-equilibrium payoff is equal to the maximum between the informative-equilibrium matching payoff and the reservation payoff $\mu$. Therefore an increase in $p$ weakly increases the receiver's informative-equilibrium payoff, and, by part 1 of this proposition, strictly so if the informative equilibrium following the increase in $p$ is influential because then the equilibrium payoff is equal to the matching payoff that lies strictly above the reservation payoff.
\end{enumerate}

\pagebreak
\addtocontents{toc}{\protect\setcounter{tocdepth}{1}}
\bibliographystyle{chicago}
\bibliography{main}

\pagebreak

\section{Online Appendix}

\subsection{A continuum of sender types}
\label{sec:continuumsenderapp}

In this Appendix, I consider an extension of my main model where the sender's type $t$ is drawn according to a continuous distribution $Q(t)$ that admits density $q(t)$ and support $[0,1]$. Assume that the signal distribution $F(\cdot|t,s)$ is continuously differentiable in $t$ for each state $s$. Here, \cref{asp:lr2} and the definition of the sender's payoff in my main model are generalized as follows:

\begin{asp*}[A lower type's experiment is noisier]\label{asp:lr4}
The following hold.
\begin{itemize}\itemsep0em
\item[\textnormal{(a)}] $\ubar \ell^t$ is strictly decreasing in $t$ and $\bar \ell^t$ is strictly increasing in $t$.
\item[\textnormal{(b)}] For each state $s$, each type $t'$, and each $\ell \in (\ubar \ell^{t'}, \bar \ell^{t'})$,
\begin{align*}
\frac{f(\ell|t',s)}{F(\ell|t',s)} > \frac{f(\ell|t,s)}{F(\ell|t,s)} \quad \text{ and } \quad \frac{f(\ell|t',s)}{1-F(\ell|t',s)} > \frac{f(\ell|t,s)}{1-F(\ell|t,s)}
\end{align*}
for all types $t > t'$.
\end{itemize}
\end{asp*}

The sender's payoff $r(m,s,\hat \gs)$ given report $m$, revelation of state $s$, and market conjecture $\hat \gs$ is
\begin{align*}
r(m,s, \hat \gs) := \int_0^1 q(t|m, s, \hat \gs) v(t) \diff t,
\end{align*}
where $q(\cdot|m,s,\hat \gs)$ denotes the market's posterior density of the sender's type given $(m,s,\hat \gs)$.

My main results extend with natural modifications. The payoff $r^*(m,s,\gb)$ in any informative equilibrium $\gb$ identified in \cref{prop:existence} is now equal to $\int_0^1 q(t|m,s,\gb) v(t) \diff t$, where the posterior density in this equilibrium satisfies
\begin{align*}
q(t|0,s,\gb) = 
\frac
{q(t) H_{\mu, F}(\gb|t,s)}
{\int_0^1 q(x) H_{\mu, F}(\gb|x,s) \diff x} \quad \text{ and } \quad q(t|1,s,\gb) = 
\frac
{q(t) (1-H_{\mu, F}(\gb|t,s))}
{\int_0^1 q(x) (1-H_{\mu, F}(\gb|x,s)) \diff x}
\end{align*}
by Bayes' rule for each state $s$, and with crossing beliefs $\gb^\dagger_0$ and $\gb^\dagger_1$ that are now defined as follows: $\gb^\dagger_0$ is the unique $\gb \in (0,1)$ satisfying
\begin{align}\label{eq:gb0daggergeneral}
\int_0^1 q(t|0,0,\gb) v(t) \diff t = \int_0^1 q(t|1,0,\gb) v(t) \diff t,
\end{align}
and 
$\gb^\dagger_1$ is the unique $\gb \in (0,1)$ satisfying
\begin{align}\label{eq:gb1daggergeneral}
\int_0^1 q(t|1,1,\gb) v(t) \diff t = \int_0^1 q(t|0,1,\gb) v(t) \diff t.
\end{align}
As in my main model, $\gb^\dagger_0$ (resp., $\gb^\dagger_1$) is the state belief given which, conditional on state 0 (resp., 1), report 0 would induce the same sender's payoff as report 1 does if this state belief were the belief cutoff in the sender's equilibrium strategy profile. At the end of this section, I formally show that these crossing beliefs are well-defined. The analogy of the high type being sufficiently informative in  \cref{thm:complementarity} is that for some $T \in (0,1)$, the experiments of sender types above $T$ are sufficiently informative relative to $\mu$ and the experiments of the other sender types: there exists $\gd_{\mu, (F^t)_{t<T}} > 0$ such that if $\max_{t \in [T,1]} \max_{s \in S} \diff (F(\cdot|t,s), F^*(\cdot|s)) < \gd_{\mu, (F^t)_{t<T}}$. The local $L^1$ convergence in \cref{thm:highermuhigherpayoff} is amended analogously.


In proving \cref{thm:complementarity}, the sequence $(F^k)_{k=0}^\infty$ of experiment profiles is now one where, along the sequence, the experiment of each type $t < T$ is constant while the experiment of each type $t \ge T$ tends to the perfectly informative experiment $F^{t*}$. Finally, the analog of  \eqref{eq:formalize} is
\begin{align*}
\bE^{\gb_{\mu,p,F^k}}[u(m,s)|t \ge T] &> \mu + \frac{Q(T)}{1-Q(T)} \!\left( \mu - \bE^{\gb_{\mu,p,F^k}}[u(m,s)|t < T]
\right)\!.
\end{align*}
The proof of \cref{thm:complementarity} is amended in the natural way.

It remains to show that the crossing beliefs are well-defined. Consider first $\gb^\dagger_0$. Note that the left side of \eqref{eq:gb0daggergeneral} is strictly decreasing in $\gb$. To prove this, because $v$ is strictly increasing, it suffices to show that for any $\gb, \gb' \in [0,1]$ with $\gb'>\gb$, the likelihood ratio $q(t|0,0,\gb')/q(t|0,0,\gb)$ is strictly decreasing in $t$. This latter claim holds because
\begin{align}\nonumber
&~ \frac{\partial}{\partial t} \!\left(
\frac{q(t|0,0,\gb')}
{q(t|0,0,\gb)}
\right)\!\\ \nonumber
=&~  \frac{\partial}{\partial t} \!\left(
\frac{H_{\mu, F}(\gb'|t,0)}
{H_{\mu, F}(\gb|t,0)}
\right)\! \\[0.25em] \label{eq:derivateHt}
=&~ 
\dfrac
{H_{\mu, F}(\gb|t,0) \frac{\partial}{\partial t} H_{\mu, F}(\gb'|t,0)
-
H_{\mu, F}(\gb'|t,0) \frac{\partial}{\partial t} H_{\mu, F}(\gb|t,0)
}
{H_{\mu, F}(\gb|t,0)^2} < 0,
\end{align}
where the inequality uses the fact that $H_{\mu, F}(\gb|t,0) \le H_{\mu, F}(\gb'|t,0)$ and that, by \cref{asp:lr4}(b),
\begin{align*}
\frac{\partial}{\partial t} H_{\mu, F}(\gb'|t,0) - \frac{\partial}{\partial t} H_{\mu, F}(\gb|t,0) = \int_{\frac{1-\mu}{\mu} \frac{\gb}{1-\gb}}^{\frac{1-\mu}{\mu} \frac{\gb'}{1-\gb'}} 
f(\ell|t,0) \diff \ell 
< 0.
\end{align*}
The right side of \eqref{eq:gb0daggergeneral} can similarly be shown to be strictly increasing in $\gb$. By \cref{asp:lr4}(a), $\ubar \gb^1 < \ubar \gb^t < \bar \gb^t < \bar \gb^1$ for every type $t \in [0,1)$. Because $v$ is strictly increasing, there exists $t^* \in (0,1)$ such that $v(t^*) > \bE^Q[v(t)]$, where the expectation is taken over $t$ with respect to the distribution $Q$. Moreover, given any such $t^*$, because the left side of \eqref{eq:gb0daggergeneral} is strictly decreasing in $\gb$, there exists $\ve_{t^*} \in (0, \bar \gb^1 - \ubar \gb^1)$ such that 
\begin{align}\label{eq:biggertstar}
\lim_{\gb \downarrow \ubar \gb^1+\ve_{t^*}} \int_0^1 q(t|0,0,\gb, \tau) v(t) \diff t > v(t^*).
\end{align}
Fix one such $t^*$ and $\ve_{t^*}$. Then
\begin{align*}
\lim_{\gb \downarrow \ubar \gb^1} \int_0^1 q(t|0,0,\gb) v(t) \diff t &> \lim_{\gb \downarrow \ubar \gb^1+\ve_{t^*}} \int_0^1 q(t|0,0,\gb) v(t) \diff t \\
&> v(t^*)\\ 
&> \bE^Q[v(t)] \\
&= \lim_{\gb \downarrow \ubar \gb^1} \int_0^1 q(t|1,0,\gb) v(t) \diff t,
\end{align*}
where the first line follows because the left side of \eqref{eq:gb0daggergeneral} is strictly decreasing in $\gb$, the second line follows uses \eqref{eq:biggertstar}, the third line follows by the choice of $t^*$, and the fourth line follows because all sender types report 1 given cutoff $\gb=\ubar \gb^1$. Similarly, there exists $\hat \ve_{t^*} \in (0, \bar \gb^1 - \ubar \gb^1)$ such that 
\begin{align*}
\lim_{\gb \uparrow \bar \gb^1} \int_0^1 q(t|0,0,\gb) v(t) \diff t &= \bE^Q[v(t)] \\
&< v(t^*) \\
&< \lim_{\gb \uparrow \bar \gb^1 - \hat \ve_{t^*}} \int_0^1 q(t|1,0,\gb) v(t) \diff t \\
&< \lim_{\gb \uparrow \bar \gb^1} \int_0^1 q(t|1,0,\gb) v(t) \diff t.
\end{align*}
As a result, there exists a unique $\gb^\dagger_0 \in (\ubar \gb^1, \bar \gb^1)$ such that \eqref{eq:gb0daggergeneral} holds with $\gb = \gb^\dagger_0$. By analogous derivations, there exists a unique $\gb^\dagger_1 \in (\ubar \gb^1, \bar \gb^1)$ such that \eqref{eq:gb1daggergeneral} holds with $\gb = \gb^\dagger_1$.

\subsection{Binary private communication}
\label{sec:binaryprivate}

In this Appendix, I show that in the private-communication extension, it is without loss of generality to focus on binary messages. In this extension, given receiver's action $a$, revelation of state $s$, and the market's conjecture $\hat \gs = (\hat \gs^h, \hat \gs^l)$ of the sender's strategy as well as its conjecture $\hat \tau$ of the receiver's strategy, the probability that the market assigns to the sender being type $t$ is $\rho(t|a, s, \hat \gs, \hat \tau)$ and each sender type's payoff as her expected market value is $r(a, s,\hat \gs, \hat \tau) := \rho(h|a, s,\hat \gs, \hat \tau)$, given the normalization $v(h)=1$ and $v(l)=0$.

\begin{prop*}[Binary messages]\label{prop:binaryc}
If an influential equilibrium $(\gs, \tau)$ exists, then there exists an influential equilibrium $(\gs', \tau')$ in which $M=S$, both messages $0$ and $1$ are sent on path, and the receiver matches his action with the sender's message. The two equilibria are outcome-equivalent: they induce the same distribution over the sender's payoffs \eqref{eq:reppayoff} and the same distribution over the receiver's actions.
\end{prop*}

\begin{proof}[Proof of \cref{prop:binaryc}]
Fix an influential equilibrium $(\gs, \tau)$. Let $\bP^{(\gs,\tau)}$ denote the probability measure over outcomes in this equilibrium. By \cref{def:vcomm}, in this equilibrium, there exists a message $m_0$ sent on path given which $\bP^{(\gs,\tau)}[s=0|m_0] > \frac 1 2$ so that the receiver has a strict incentive to take action 0. Because $\bP^{(\gs,\tau)}[s=1] = \mu \in [\frac 1 2, 1)$, by the law of iterated expectations on the receiver's state belief, in this equilibrium there exists another message $m_1$ sent on path given which $\bP^{(\gs,\tau)}[s=1|m_1] > \frac 1 2$ so that the receiver has a strict incentive to take action 1. 

I proceed via several claims.

\begin{claim*}\label{claim:binaryc1}
There exists an influential equilibrium $(\hat \gs, \tau)$ so that in the equilibrium, for each state $s \in S$, there exists a unique message, denoted by $s$, sent on path given which the receiver has a strict incentive to take action $s$, and the two equilibria $(\gs, \tau)$ and $(\hat \gs, \tau)$ induce the same sender's payoff \eqref{eq:reppayoff} given each action and state revelation as well as the same distribution over the receiver's actions. Moreover, in the equilibrium $(\hat \gs, \tau)$, upon receiving any message $m \notin S$ that is sent on path, if ever, the receiver totally mixes between the two actions.
\end{claim*}

\begin{proof}[Proof of \cref{claim:binaryc1}]
Let $M^\gs$ be the set of messages that are sent on path in the equilibrium $(\gs, \tau)$. If this equilibrium satisfies the properties of $(\hat \gs, \tau)$ stated in the claim, then the claim immediately follows. Suppose then that it does not. Then there exists an action $a^* \in S$ such that for some $m, m' \in M^\gs$, $m \neq m'$, the receiver takes action $a^*$ with probability one upon receiving either message $m$ or $m'$ in the equilibrium, and he has a strict incentive to take this action upon receiving at least one of these two messages; one such message must exist because $m_{a^*}$ is one such message. Therefore, the following two inequalities hold:
\begin{align}\label{eq:bigger}
\bP^{(\gs, \tau)}[s=a^*|m] 
&= \frac{\bP[s=a^*] \ \bE^p[ \int_{Y(t)} \gs^t(m|y) \diff G(y|t,a^*) ]}{
\bE^{\mu,p}[ \int_{Y(t)} \gs^t(m|y) \diff G(y|t,s) ] 
]
} \ge \frac 1 2,\\
\bP^{(\gs, \tau)}[s=a^*|m'] 
&= \frac{\bP[s=a^*] \ \bE^p[ \int_{Y(t)} \gs^t(m'|y) \diff G(y|t,a^*) ]}{
\bE^{\mu,p}[ \int_{Y(t)} \gs^t(m'|y) \diff G(y|t,s) ] 
} \ge \frac 1 2,\label{eq:bigger2}
\end{align}
with at least one of these inequalities being strict, where the expectation $\bE^{\mu,p}[\cdot]$ is taken with respect to both the prior state distribution and the prior type distribution over $s$ and $t$. Consider another strategy profile $(\gs^\dagger, \tau)$, where $\gs^\dagger$ is almost identical to $\gs$ except that $m'$ is relabeled as $m$, and let $M^{\gs^\dagger} := M^\gs \setminus \{ m' \}$ be the set of messages that are sent on path given this strategy profile. 

To complete the proof, I first show that $(\gs^\dagger, \tau)$ is an influential equilibrium. Write $M^{\gs^\dagger}$ and $\bP^{(\gs^\dagger,\tau)}$ as the counterpart of $M^\gs$ and $\bP^{(\gs,\tau)}$ above. Observe that $\tau$ is a best reply to $\gs^\dagger$ and captures the receiver's strict incentive to take action $a^*$ upon receiving message $m$ when conjecturing $\gs^\dagger$ because $\tau$ is a best reply to $\gs$ by assumption, because $\bP^{(\gs^\dagger, \tau)}[s=a|\tilde m] = \bP^{(\gs, \tau)}[s=a|\tilde m]$ for any action $a$ and any message $\tilde m \in M^{\gs^\dagger} \setminus \{m\}$, and because
\begin{align*}
\bP^{(\gs^\dagger,\tau)}[s=a^*|m] 
&= 
\frac{\bP[s=a^*] \ \bE^p[ \int_{Y(t)} \gs^{\dagger, t}(m|y) \diff G(y|t,a^*) ]}{
\bE^{\mu, p}[ \int_{Y(t)} \gs^{\dagger, t}(m|y) \diff G(y|t,s) ] 
}\\
&= 
\frac{\bP[s=a^*] \ \bE^p[ \int_{Y(t)} \gs^t(m |y) + \gs^t(m' |y) \diff G(y|t,a^*) ]}
{
\bE^{\mu,p}[ \int_{Y(t)} \gs^t(m |y) + \gs^t(m' |y) \diff G(y|t,s) ] 
}\\
&> \frac 1 2,
\end{align*}
where the inequality follows by \eqref{eq:bigger} and \eqref{eq:bigger2}, and this inequality is strict because either \eqref{eq:bigger} or \eqref{eq:bigger2} is strict. On the other hand, $\gs^{\dagger, t}$ is each sender type $t$'s best reply given the market's conjecture $(\gs^\dagger, \tau)$ and her conjecture $\tau$ of the receiver's strategy. This is because her best reply problem given the market's conjecture $(\gs^\dagger, \tau)$ and her conjecture $\tau$ of the receiver's strategy is identical to her best reply problem given the market's conjecture $(\gs, \tau)$ and her conjecture $\tau$ of the receiver's strategy.
\end{proof}



Next, fix an influential equilibrium, labeled again as $(\gs, \tau)$ for simplicity, and in view of \cref{claim:binaryc1}, suppose that in this equilibrium, for each state $s$, there exists a unique message, denoted by $s$, sent on path given which the receiver takes action $s$ with probability one; moreover, in this equilibrium, upon receiving message $s$, the receiver's incentive to take action $s$ is strict. Let $\neg s$ denote the state other than $s$. Call message $s$ a report that the state is $s$. To simplify notation, in the rest of this proof, I write the payoff $r(a,s,\gs,\tau)$ simply as $r(a,s,\gs)$ for each $(a,s)$. In this equilibrium, given state belief $b$, the sender's payoff from reporting 0 is \eqref{eq:0payoff} and her payoff from reporting 1 is \eqref{eq:1payoff}. Therefore she prefers reporting 1 to reporting 0 if and only if \eqref{eq:fixedpointbeta0} holds.

Next, I show that any message given which the receiver plays a totally mixed action must be sent with probability zero in the equilibrium $(\gs, \tau)$. Suppose that in the equilibrium there exists a message $m^\dagger \notin S$ given which the receiver takes action 0 with some probability $\ga \in (0,1)$ and takes action 1 with complementary probability. 

\begin{claim*}\label{claim:probzero}
In the equilibrium $(\gs, \tau)$, the message $m^\dagger$ is sent with probability zero.
\end{claim*}

\begin{proof}[Proof of \cref{claim:probzero}]
In the equilibrium,  each sender type's payoff from sending message $m^\dagger$, given state belief $b$, the market's conjecture $(\gs, \tau)$ and her conjecture $\tau$ of the receiver's strategy, is
\begin{align}\nonumber
&~ b [\ga r(0,1,\gs) + (1-\ga) r(1,1,\gs)] 
+ 
(1-b) [ \ga r(0,0,\gs) + (1-\ga) r(1,0,\gs)] \\ \label{eq:mdaggerpayoff}
=&~
\ga [b r(0,1,\gs) + (1-b) r(0,0,\gs) ] + (1-\ga) [ b r(1,1,\gs) + (1-b) r(1,0,\gs)],
\end{align}
which is a weighted average of \eqref{eq:0payoff} and \eqref{eq:1payoff}. Consequently, she optimally sends message $m^\dagger$ only if given state belief $b$, \eqref{eq:0payoff} and \eqref{eq:1payoff} are equal so that she is indifferent between sending message 0 and message 1:
\begin{align}\label{eq:fixedpointbeta0eq}
b [ r(1,1,\gs) - r(0,1,\gs)] = (1 - b) [ r(0,0,\gs) - r(1,0,\gs)]. 
\end{align}
Given \eqref{eq:correctreport}, there exists a unique $\hat b \in \bR$ such that \eqref{eq:fixedpointbeta0eq} holds if and only if her state belief $b$ is equal to $\hat b$, which happens with zero probability.
\end{proof}

To complete the proof, construct a new equilibrium $(\gs', \tau')$ from the influential equilibrium $(\gs, \tau)$ by relabeling all messages given which the receiver plays a totally mixed action as 1. Given the strategy profile $(\gs', \tau')$, only messages 0 and 1 are sent on path. By construction, $\gs'$ is a best reply to $\tau'$ because given the market's conjecture $\gs$ and the receiver's strategy $\tau$, $\gs^t$ is each sender type $t$'s best reply to $\tau$ and, when sending any message given which the receiver plays a totally mixed action, she is indifferent between inducing action 0 and action 1 by \cref{claim:probzero}. On the other hand, the receiver's strategy $\tau'$ is a strict best reply to $\gs'$. He has a strict incentive to match his action with report 0 when conjecturing $\gs'$ because he has a strict incentive to do so when conjecturing $\gs$. Similarly, the receiver has a strict incentive to match his action with report 1 when conjecturing $\gs'$ because first, he has a strict incentive to do so when conjecturing $\gs$ and second, he perceives that the event of a message being sent given which he plays a totally mixed action under $\gs$ happens with zero probability by \cref{claim:probzero}. Therefore $(\gs', \tau')$ is an equilibrium and it is influential. Finally, because the event of a message being sent given which the receiver plays a totally mixed action under $\gs$ happens with zero probability by \cref{claim:probzero}, the two influential equilibria $(\gs, \tau)$ and $(\gs', \tau')$ induce the same distribution over the sender's payoffs as well as the same distribution over the receiver's actions, as desired.
\end{proof}

\end{document}